\documentclass[aps,pre,twocolumn,showpacs,%
preprintnumbers,amssymb]{revtex4-1}
\usepackage{graphicx}
\usepackage[colorlinks=true,linkcolor=blue,%
pagecolor=blue,citecolor=blue,%
urlcolor=blue,anchorcolor=black,%
bookmarksnumbered=true,%
bookmarksopen=true]{hyperref}
\usepackage{dcolumn}
\usepackage{bm}
\renewcommand{\frac}[2]{\displaystyle{#1 \over #2}}
\begin{document}
\title{Ionization equation of state for the dusty 
plasma including the effect of ion--atom 
collisions}
\author{D.~I.~Zhukhovitskii} \email{dmr@ihed.ras.ru}
\affiliation{Joint Institute of High Temperatures, Russian Academy of 
Sciences, Izhorskaya 13, Bd.~2, 125412 Moscow, Russia}
\affiliation{Moscow Institute of Physics and Technology, Institutsky 
lane 9, Dolgoprudny, Moscow region, 141700 Russia }
\date{\today}
\begin{abstract}
The ionization equation of state (IEOS) for a cloud of 
the dust particles in the low-pressure gas discharge under 
microgravity conditions is proposed. IEOS relates pairs 
of the parameters specific for the charged components of 
dusty plasma. It is based on the modified collision 
enhance collection model adapted for the Wigner--Seitz 
cell model of the dust cloud. This model takes into 
account the effect of ion--atom collisions on the ion 
current to the dust particles and assumes that the 
screening length for the ion--particle interaction is of the 
same order of magnitude as the radius of the 
Wigner--Seitz cell. Included effect leads to a noticeable 
decrease of the particle charge as compared to the 
previously developed IEOS based on the orbital motion 
limited model. Assuming that the Havnes parameter of 
the dusty plasma is moderate one can reproduce the dust 
particle number density measured in experiments and, in 
particular, its dependence on the gas pressure. Although 
IEOS includes no fitting parameters, it can ensure a 
satisfactory precision in a wide range of dusty plasma 
parameters. Based on the developed IEOS, the threshold 
relation between the dusty plasma parameters for onset 
of the lane formation in binary dusty plasmas is deduced.
\end{abstract}
\maketitle
\section{\label{s1}INTRODUCTION}

Low-temperature plasmas that contain dust particles typically in the 
range from $0.01$
 to $1000\,\mu {\mbox{m}}$
 are termed dusty (or complex) plasmas \cite{1,2,3,5,6,9}. Laboratory 
dusty plasmas are generated to study fundamental processes in the 
strong coupling regime on the kinetic level by the observation of 
individual microparticles. Due to the high electron mobility, particles 
acquire a considerable negative electric charge. Because of the 
Coulomb repulsion, they can form extended clouds. In the 
ground-based experiments, gravity is one of the crucial forces that 
define the properties of a dust cloud. Under microgravity conditions, 
e.g., on the International Space Station (ISS) 
\cite{10,15,16,17,18,019,19} or in parabolic flights 
\cite{10,11,12,13,14}, the particles can form almost homogenous 
three-dimensional clouds in the bulk of the low-pressure gas 
discharge. In addition, due to the large particle charge, the Coulomb 
coupling parameter of the particle subsystem is great, so that such 
subsystem can form an analog of condensed state of matter, i.e., 
three-dimensional liquid or solid.

One of the basic objectives in this field is the investigations of 
correlations between the governing parameters of dusty plasmas, in 
particular, the spatial distribution of the local particle number density 
in a stationary dust cloud $n_d$ under different conditions of the gas 
discharge. Thus, the dust distribution under the conditions of PKE 
(plasma crystal experiment) chamber was investigated numerically 
\cite{107}; equation of state for 2D liquid dusty plasmas was 
obtained in \cite{106}; the dust distribution in the sheath under the 
conditions of PK-3 Plus chamber is studied in \cite{108}. In the 
works \cite{22,35,86}, the particle distribution in a 
quasi-homogeneous region of the dust cloud apart from the void was 
modeled by construction of the ionization equation of state (IEOS). 
IEOS is a relation between a pair of the parameters specific for the 
charged components of complex plasma containing a cloud of the 
dust particles. Such parameters are the electron, the ion, and the 
particle number density, and the particle potential (related to its 
charge). A complete set of IEOS's makes it possible to calculate all 
plasma state parameters provided that a single one is known. This 
makes IEOS similar to the common equation of state. The IEOS 
proposed in \cite{22,35,86} employs the balance equation for the 
main forces acting on a dust particle, the quasineutrality equation, and 
the particle charge equation. The latter is based on the orbital motion 
limited approximation \cite{55} (OML), which was shown to 
underestimate substantially the ion flux toward the particle due to 
disregard of the ion--atom collisions \cite{43}. This leads to 
overestimation of the particle charge and, correspondingly, to 
underestimation of $n_d$ \cite{22,35,86} as compared to the 
experiment. In addition, the estimated electron and ion number 
densities seem to be overestimated by more than an order of 
magnitude. The dependence of $n_d$ on the gas pressure observed in 
experiment is not reproduced by such IEOS, even if the dependence 
of the ion mean free path on the local particle number density is 
properly taken into account \cite{86}.

To modify our approach, we adopt the expression for the ion flux 
\cite{43}, which was obtained for the case of a solitary particle in 
plasma, and change it for the case of a dust cloud. We show that a 
relevant model for the cloud is the Wigner--Seitz cell model, in which 
the screening length is of the same order of magnitude as the cell 
radius. With this screening length, one can obtain a correct expression 
for the ion flux to the particle and derive the particle charge equation. 
In the modification of IEOS proposed in this work, we use this 
expression instead of that based on the OML approach \cite{86}. The 
IEOS obtained in this study makes it possible to attain a good 
correlation between the magnitudes of all plasma parameters. In 
particular, it ensures a correct dependence of $n_d$ on the gas 
pressure with due regard for the dependence of the electron number 
density on the pressure. At the same time, this IEOS is free from 
fitting parameters. Nevertheless, it is valid in a wide range of dusty 
plasma parameters and can ensure a sufficient precision. In particular, 
this provides an interpretation for the decreasing dependence of 
$n_d$ on the pressure observed in the experiment \cite{43}.

We use this new modification of IEOS to estimate the threshold 
relation between plasma parameters corresponding to the onset of 
lane formation in the binary complex plasmas observed in 
experiments \cite{68,63,92}. This effect takes place if small particles 
are injected in a stationary cloud of large particles. Under the 
experimental conditions, the latter is typically a dust crystal. We 
assume that the lane formation is a manifestation of the crystal 
spallation entailed by the injection of small particles. Hence, the work 
of the driving force acting on the small particles must be greater than 
the work against the pressure of the particles that form a dust cloud. 
We calculate the driving force in the same way as in \cite{86}, 
however, we apply the new modification of IEOS to attain a 
correlation with the magnitudes of experimental parameters.

Proposed IEOS is significant for both understanding the properties of 
dusty plasmas and planning the future experiments. Thus, based on a 
standard simulation of the discharge without particles and the 
developed IEOS, one can estimate the parameters of complex plasma 
for the Ekoplasma project, which is a Russian--German cooperation 
building the future laboratory for the investigation of complex 
plasmas under microgravity conditions on the ISS \cite{110}. Such 
calculations will enable optimization of the conditions of forthcoming 
experiments.

The paper is organized as follows. In Sec.~\ref{s2}, the screening 
length for a dense 3D cloud of particles in a low-pressure discharge 
plasma is estimated on the basis of the Wigner--Seitz cell model. In 
Sec.~\ref{s3}, the effect of the ion--atom collisions on the ion current 
to the particle is estimated for the treated system, and the equation 
defining the particle charge is derived. The IEOS taking into account 
this effect is obtained and analyzed in Sec.~\ref{s4}. In 
Sec.~\ref{s5}, the calculation results using obtained IEOS are 
compared with available experimental data. In Sec.~\ref{s6}, the 
proposed IEOS is applied to the calculation of the threshold for onset 
of the lane formation in binary dusty plasmas. The results of this 
study are summarized in Sec.~\ref{s7}.

\section{\label{s2}SCREENING LENGTH IN THE CELL 
MODEL OF A DENSE DUST CLOUD}

Consider a stationary cloud of dust particles of the same radius in the 
low-pressure gas discharge. We will treat a ``dense'' cloud, in which 
the interparticle correlations are as strong as in the condensed state of 
matter. In fact, such cloud can be a model either for liquid or solid 
state. For this system, the Wigner--Seitz cell model will be utilized. It 
implies that dusty plasma is divided in spherical cells with the radius 
$r_d = (3/4\pi n_d )^{1/3} $. Each particle finds itself in the center of 
a spherical cell filled with the background volume charge from the 
light plasma components (electrons and ions). The cell as a whole is 
electrically neutral, i.e., the electric field at its boundary vanishes.

Under typical conditions of the low-pressure gas discharge, the 
electrons are fully thermalized and they obey the Boltzmann 
distribution. If the particle dimensionless electric potential $\Phi = 
Ze^2 /aT_e $, where $Z$
 is the particle charge in units of the electron charge, $e$
 is the elementary electric charge, $a$
 is the particle radius, and $T_e$ is the electron temperature, is 
restricted by the assumed condition $\Phi < 1$
 then the inhomogeneity in the electron spatial distribution is only 
insignificantly different from a constant in the vicinity of the particle 
and cannot screen noticeably the particle charge. In addition, note that 
typically $T_e \sim 4\;{\mbox{eV}}$
 and the electron number density $n_e \sim 3 \times 10^8 {\kern 1pt} 
{\mbox{cm}}^{ - 3} $, which means that the electron Debye 
screening length $r_{De} = (T_e /4\pi n_e e^2 )^{1/2} \sim 
0.1\;{\mbox{cm}}$
 is sufficiently large to satisfy the condition $r_d \ll r_{De} $.

The ions are far from equilibrium with the particles. The spatial 
distribution of the ion number density in the vicinity of a particle 
depends on the relations between the particle radius, the ion mean 
free path with respect to the collisions with atoms $\lambda _a $, and 
the length of dust particle screening. In the collisionless regime for a 
solitary small particle in the infinite stationary plasma, OML results 
in the buildup of ion number density around the particle and 
consequent screening length close to the Debye one \cite{1} $r_{Di} 
= (T_i /4\pi n_i e^2 )^{1/2} $, where $n_i$ is the average ion number 
density. Due to low ion temperature $T_i \sim 0.03\;{\mbox{eV}}$, 
for $n_i \sim n_e $, we obtain $r_d \sim r_{Di}$ for the small 
particles and $r_d \gg r_{Di}$ for the large ones. If the dust cloud is 
modeled by the Yukawa system then the particle electric potential is 
prescribed the Yukawa form. However in the Wigner--Seitz cell 
model, the particle potential is significantly different from the 
Yukawa one even for the Boltzmann ion number density distribution, 
albeit the particle charge can be screened (renormalized) by a thin 
layer around a particle. This is the case when the local Debye 
screening length is smaller than the particle radius \cite{103}. In our 
case, this would require $n_i > 10^{12} \,{\mbox{cm}}^{ - 3} $, 
which seems unrealistic because of the non-exponential 
nonequilibrium radial dependence of $n_i $. Apparently, one can 
neglect the Debye screening in the cell in the collisionless case. In the 
opposite case of highly collisional regime ($a \gtrsim \lambda _a $) in 
the neighborhood of a particle, $n_i$ is lower than its volume average 
(see, e.g., \cite{22}). This excludes any ion screening of the particle 
charge in this region. In the region of moderate collisions treated in 
this study the potential of a solitary particle is almost the Coulomb 
one \cite{104}, and screening is absent. Unfortunately, to the best of 
our knowledge, no calculation of the ion number density distribution 
in the Wigner--Seitz cell model is available in the literature. At the 
same time, it seems reasonable to treat the total charge background in 
the cell $e(n_i - n_e )$
 as a uniform one.

One can conclude that in the cell model, the particle charge screening 
is different from the Debye one and the ion Debye length is no 
appropriate scale for the system. In this model, screening is caused by 
the overall cell quasineutrality. Hence, the screening length must be 
of the order of the cell radius $r_d $. In what follows, we will define 
this length.

The distribution of electric potential in the cell $\varphi (r)$
 is defined by the Poisson equation
\begin{equation}
\frac{1}{{r^2 }}\frac{d}{{dr}}\left( {r^2 \frac{{d\varphi }}{{dr}}} 
\right) = 4\pi e(n_e - n_i ), \label{e1}
\end{equation}
which should be solved with the boundary conditions
\begin{equation}
\varphi (r_d ) = \varphi '(r_d ) = 0,\quad \varphi '(a) = 
\frac{{Ze}}{{a^2 }}. \label{e2}
\end{equation}
We represent the solution in the form
\begin{equation}
\varphi (r) = \frac{A}{r} + Br^2 + C, \label{e3}
\end{equation}
where $A$, $B$, and $C$
 are constants, to obtain
\begin{equation}
\varphi (r) = - \frac{{Ze}}{{r_d }}\frac{1}{{1 - (a/r_d )^3 }}\left( 
{\frac{{r_d }}{r} + \frac{1}{2}\frac{{r^2 }}{{r_d^2 }} - 
\frac{3}{2}} \right). \label{e4}
\end{equation}
Solution (\ref{e4}) is compatible with the cell quasineutrality 
condition
\begin{equation}
(n_i - n_e )\left( {1 - \frac{{a^3 }}{{r_d^3 }}} \right) - Zn_d = 0. 
\label{e5}
\end{equation}
In what follows, we will neglect the very small ratio $a^3 /r_d^3 \sim 
10^{ - 6} $.
\begin{figure}
\includegraphics[width=0.9\columnwidth]{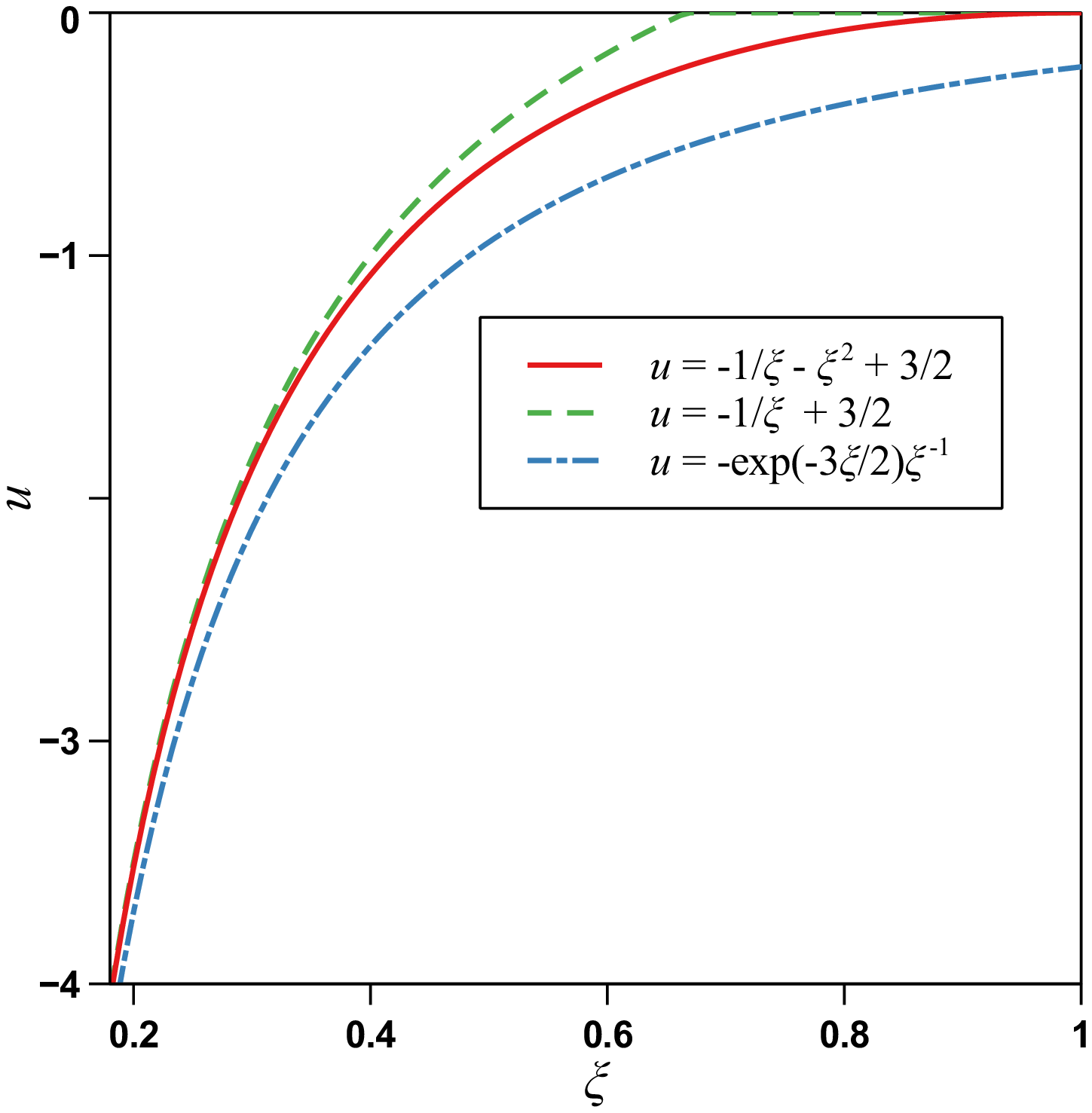}
\caption{\label{f1}Particle potential in the Wigner--Seitz cell (solid 
line), its approximation by the shifted Coulomb potential (dashed 
line), and by the Yukawa potential (dash-dotted line).}
\end{figure}

The dimensionless ion potential energy in the cell $e\varphi (r)/T_i = 
(a\tau \Phi /r_d )u(r/r_d )$, where $\tau = T_e /T_i $, $u(\xi ) = - 1/\xi 
- \xi ^2 /2 + 3/2$, and $\xi = r/r_d $, is shown in Fig.~\ref{f1}. The 
function $u(\xi )$
 can be approximated by the shifted Coulomb potential $u(\xi ) 
\simeq - 1/\xi + 3/2$
 if $\xi \le 2/3$, which is the asymptotic of $u(\xi )$
 at $\xi \ll 1$, and $u(\xi ) \equiv 0$
 if $\xi > 2/3$. The same asymptotic is characteristic of the Yukawa 
potential $ - \exp ( - 3\xi /2)\xi ^{ - 1}$ (Fig.~\ref{f1}). Note that the 
Yukawa potential approximates $u(\xi )$
 significantly worse than the shifted Coulomb potential. 
Consequently, in the cell model, the screening length can be defined 
as $r_s = (2/3)r_d $. For the clouds of large particles, $r_s$ can be 
considerably larger than $r_{Di} $. Since the region essential for the 
moment transfer from the ions to a particle is restricted by the 
condition $u \lesssim - 1$, i.e., $r \lesssim 0.45r_d $, $r_s$ could be 
defined otherwise. However, it would always be $r_s \sim r_d$ for 
any definition.

Next, we will discuss the condition of applicability of the 
Wigner--Seitz cell model for the particle subsystem. Obviously, this 
model is appropriate for a highly correlated system of particles, in 
which the displacement of particles from their equilibrium positions 
in the crystal is much smaller than the interparticle distance. Hence, 
the applicability condition can be obtained in the same way as for the 
Wigner electron crystal, which differs from the treated system in the 
charge signs of the particles and the background. If the background is 
assumed uniform then a particle oscillates in the spherical potential 
well \cite{103} $U(\delta r) = (2\pi /3)n_d Z^2 e^2 (\delta r)^2 $, 
where $\delta r$
 is the deviation of a particle from the center of the cell. If $\delta r$
 is equal to the rms deviation from the cell center then $U(\delta r) = 
M\left\langle {v^2 } \right\rangle /2 = 3T_d /2$, where $M$
 is the particle mass, $\left\langle {v^2 } \right\rangle$ is its average 
velocity, and $T_d$ is the dust particle kinetic temperature (the 
Boltzmann constant is set to unity). We require the amplitude of the 
particle oscillations to be much smaller than the cell radius $r_d $, 
$(\delta r/r_d )^2 \ll 1$
 to obtain the condition \cite{94}
\begin{equation}
\Gamma = 3\left( {\frac{{r_d }}{{\delta r}}} \right)^2 \gg 1, 
\label{e5001}
\end{equation}
where $\Gamma = Z^2 e^2 /r_d T_d$ is the Coulomb coupling 
parameter. Equation (\ref{e5001}) is the condition of the model 
applicability. From (\ref{e5001}), an important conclusion follows 
that the Wigner--Seitz cell model is a model of strongly coupled 
dusty plasma \cite{103}. Under typical experimental conditions, 
$\Gamma \sim 100$, which justifies the use of this model.

As the ion--particle interaction is concerned, we note that the 
Wigner-Seitz cell model of dusty plasma implies that the volume 
screening of the particle charge by the uniform charged background is 
stronger than the particle screening by polarization of the background 
in the vicinity of a particle, as in the case of the Debye screening 
\cite{103,47}.

\section{\label{s3}PARTICLE CHARGE EQUATION}

The stationary particle charge $Z$
 is defined by the balance between the electron and ion currents to the 
particle. Since in the low-pressure RF discharge, the electrons are 
thermalized and they obey the Boltzmann distribution, the electron 
current is $j_ -  = - \pi a^2 en_e v_{Te} \exp ( - \Phi )$, where 
$v_{Te} = \left( {8T_e /\pi m_e } \right)^{1/2}$ is the electron 
thermal velocity and $m_e$ is the electron mass.

In a number of studies, it was pointed out that even in the case $r_s < 
\lambda _a$ (low-collision plasma), the OML approximation seems 
to underestimate the ion current $j_ +$ \cite{1}. The ion--atom 
collisions in a deep potential well of a particle, although rare, reduce 
the ion energy and its angular moment considerably. Hence, the 
probability that the ion trajectory can intersect the particle surface 
increases sharply. Slow ions can be created also by the atom 
ionization process that occurs, in particular, in the vicinity of 
particles. An approach to account for the ion current enhancement 
was proposed in \cite{105}. The most convenient form of the 
expression for the ion current incorporates the effect of ion--atom 
collisions and ionization. In \cite{43}, the collision enhance 
collection model (CEC) was formulated, which interpolates the ion 
current between the cases of different ratios of the plasma length 
parameters,
\begin{equation}
j_ +  = \pi a^2 en_i v_{Ti} \left[ {1 + \tau \Phi + 2.8\frac{{r_{Di}^3 
}}{{\lambda _a a^2 }}\ln ^3 \left( {1 + \frac{{a\tau \Phi }}{{r_{Di} 
}}} \right)} \right], \label{e6}
\end{equation}
where $v_{Ti} = \left( {8T_i /\pi m_i } \right)^{1/2}$ is the ion 
thermal velocity, $m_i$ is the ion mass, $\tau = T_e /T_i $, and the 
factor $2.8$
 accounts for the ionization in the vicinity of a particle. In contrast to 
\cite{43}, we will use the screening length for the cell model $r_s = 
(2/3)r_d$ rather than the Debye length $r_{Di} $. It will be shown 
below that for $a > 10^{ - 4} \,{\mbox{cm}}$, $\Phi < 0.4$. Since 
under typical experimental conditions, $T_e \sim 4\,\,{\mbox{eV}}$, 
$T_i \sim 0.03\,\,{\mbox{eV}}$, and $\tau \sim 10^2 $, one can 
assume that $a\tau \Phi /r_s \lesssim 1$
 and rewrite (\ref{e6}) as
\begin{equation}
j_ +  = \pi a^2 en_i v_{Ti} \left( {1 + \tau \Phi + 
2.8\frac{a}{{\lambda _a }}\tau ^3 \Phi ^3 } \right). \label{e7}
\end{equation}
Note that $r_s$ is canceled in (\ref{e7}). Therefore, the ion current 
(\ref{e7}) is independent of a concrete definition of $r_s $; it is only 
essential that $r_s \sim r_d $. Equation~(\ref{e7}) is similar to that 
proposed in \cite{105}.

As compared to the OML approximation, MCEC includes the third 
term in parenthesis on the r.h.s. The latter dominates, i.e., $j_ +$ is 
enhanced as compared to the OML approximation if the ratio of the 
third to the second term $2.8(a/\lambda _a )\tau ^2 \Phi ^2 > 1$
 or $\lambda _a /a < 2.8\tau ^2 \Phi ^2 \sim 10^3 $. Since typically 
$r_s /a \sim r_d /a \sim 10^2 $, this means that $\lambda _a /r_s < 
10$. Such condition is always satisfied for treated complex plasma. 
Consequently, only the condition $\lambda _a /r_s > 10$
 would be sufficient to treat collisionless complex plasma. The same 
conclusion can be found in \cite{67}. Thus, the overall result of the 
ion current enhancement is the particle charge reduction. Note that 
Eq.~(\ref{e7}) can be valid even in the case of a solitary particle in 
plasma provided that the condition $a\tau \Phi /r_s \lesssim 1$
 is satisfied, where, by and large, the screening length $r_s$ does not 
coincide with $r_{Di} $.

Thus, the ion current can be written in the form $j_ +  = 2.8\pi en_i 
v_{Ti} a^3 \tau ^3 \Phi ^3 /\lambda _a $. In what follows, this 
expression will be referred to as the modified collision enhance 
collection model (MCEC). The equation $j_ -  + j_ +  = 0$
 is then equivalent to
\begin{equation}
\theta \Phi ^3 e^\Phi  = \frac{{n_e^* }}{{n_i^* }}, \label{e8}
\end{equation}
which defines the particle charge $Z = aT_e \Phi /e^2 $. Here,
\begin{equation}
\theta = 2.8\tau ^2 \frac{a}{{\lambda _a }}\left( {\frac{{T_e m_e 
}}{{T_i m_i }}} \right)^{1/2} \label{e9}
\end{equation}
is a single parameter that defines the treated system; $n_e^* = (e^2 
\lambda _a^3 /aT_e )n_e$ and $n_i^* = (e^2 \lambda _a^3 /aT_e 
)n_i$ are the electron and ion dimensionless number densities, 
respectively. The particle potential (charge) equation differs from that 
used in recent studies \cite{22,35,86} in the definition of $\theta$ and 
in the power of $\Phi$ on the l.h.s.\ of (\ref{e8}).

\section{\label{s4}IONIZATION EQUATION OF STATE FOR 
THE DUST CLOUD}

Under microgravity conditions, a dust particle is subject to the 
electrostatic force, the ion drag force from the ions scattering on the 
dust particles, and the neutral drag force due to collisions of the atoms 
against the moving particles. For a stationary cloud, the latter force 
vanishes. The electrostatic force per unit volume is ${\bf{f}}_{ed} = 
- Zen_d {\bf{E}}$, where ${\bf{E}} = (T_e /e)\bm{\nabla}\ln n_e$ 
is the ambipolar electric field and the ion drag force is ${\bf{f}}_{id} 
= (3/8)(4\pi n_d /3)^{1/3} n_i \lambda e{\bf{E}}$
 \cite{22,35}. Here, $\lambda$ is the ion mean free path with respect 
to collisions both with the atoms and with the particles, in contrast to 
the ion mean free path in pure plasma without particles $\lambda _a 
$. $\lambda$ is calculated using a simple interpolation \cite{86}
\begin{equation}
\lambda = \lambda _a \left( {1 + \frac{3}{{8\rho }}} \right)^{ - 1} , 
\label{e10}
\end{equation}
where $\rho = r_d /\lambda _a $. Thus, the force balance equation 
${\bf{f}}_{ed} + {\bf{f}}_{id} = 0$
 yields \cite{86}
\begin{equation}
\frac{\pi }{2}\rho ^2 n_i^* = \Phi \left( {1 + \frac{3}{{8\rho }}} 
\right). \label{e11}
\end{equation}
Equation (\ref{e11}) along with the particle charge equation 
(\ref{e8}) and the quasineutrality condition (\ref{e5}) that can be 
written in the dimensionless quantities as \cite{86}
\begin{equation}
1 - \frac{3}{{4\pi }}\frac{\Phi }{{n_i^* \rho ^3 }} = \frac{{n_e^* 
}}{{n_i^* }}, \label{e12}
\end{equation}
form a set of equations that enables one to calculate all plasma state 
parameters provided that a single one is known.

Thus, from (\ref{e8}) and (\ref{e12}), it follows that
\begin{equation}
n_i^* = \frac{3}{{4\pi }}\frac{\Phi }{{\gamma (\Phi )\rho ^3 
}}\;\;{\mbox{where}}\;\;\gamma (\Phi ) = 1 - \theta \Phi ^3 e^\Phi . 
\label{e13}
\end{equation}
Then from (\ref{e11}), we obtain the IEOS in the variables $\rho$ 
and $\Phi$ ($\rho$ defines the particle number density, $n_d = (3/4\pi 
)(\rho \lambda _a )^{ - 3} $)
\begin{equation}
\frac{{8\rho }}{3} + 1 = \frac{1}{{\gamma (\Phi )}}. \label{e14}
\end{equation}
We multiply both sides of Eq.~(\ref{e13}) by $[1 + 3(8\rho )^{ - 1} 
]^{ - 3}$ to derive
\begin{equation}
\rho ^3 \left( {1 + \frac{3}{{8\rho }}} \right)^3 = \frac{3}{{4\pi 
}}\frac{\Phi }{{\gamma (\Phi )\tilde n_i }}, \label{e1401}
\end{equation}
where $\tilde n_i = n_i^* [1 + 3(8\rho )^{ - 1} ]^{ - 3} $. The same 
operation applied to Eq.~(\ref{e11}) yields
\begin{equation}
\rho = \left( {1 + \frac{3}{{8\rho }}} \right)^{ - 1} \left( 
{\frac{2}{\pi }\frac{\Phi }{{\tilde n_i }}} \right)^{1/2} . 
\label{e1402}
\end{equation}
On substitution of $\rho$ (\ref{e1402}) into (\ref{e1401}) one can 
derive the IEOS in the variables $\tilde n_i$ and $\Phi $,
\begin{equation}
\frac{3}{8}\left( {\frac{{\pi \tilde n_i }}{{2\Phi }}} \right)^{1/2} = 
\gamma (\Phi ). \label{e15}
\end{equation}
Equation (\ref{e15}) coincides with Eq.~(7) in \cite{86}, however, 
the definition of $\gamma (\Phi )$
 is different from (\ref{e13}). Combination of (\ref{e14}) and 
(\ref{e15}) yields the IEOS's
\begin{equation}
n_i^* = \frac{{128}}{{9\pi }}\frac{{\Phi \gamma ^2 }}{{(1 - 
\gamma )^3 }},\;\;n_e^* = (1 - \gamma )n_i^* = \frac{{128}}{{9\pi 
}}\frac{{\Phi \gamma ^2 }}{{(1 - \gamma )^2 }} \label{e16}
\end{equation}
in the variables $n_i^* $, $\Phi$ and $n_e^* $, $\Phi $, respectively. 
Note that the IEOS's (\ref{e14}), (\ref{e15}), and (\ref{e16}) have a 
similarity property\cite{64}. An important property of complex 
plasma, the Havnes parameter $H \equiv Zn_p /n_e$ defining the 
re-distribution of charge between the light and heavy charge carriers, 
can be obtained from (\ref{e8}) and (\ref{e13}):
\begin{equation}
H = \frac{\gamma }{{1 - \gamma }}. \label{e17}
\end{equation}

\begin{figure}
\includegraphics[width=0.9\columnwidth]{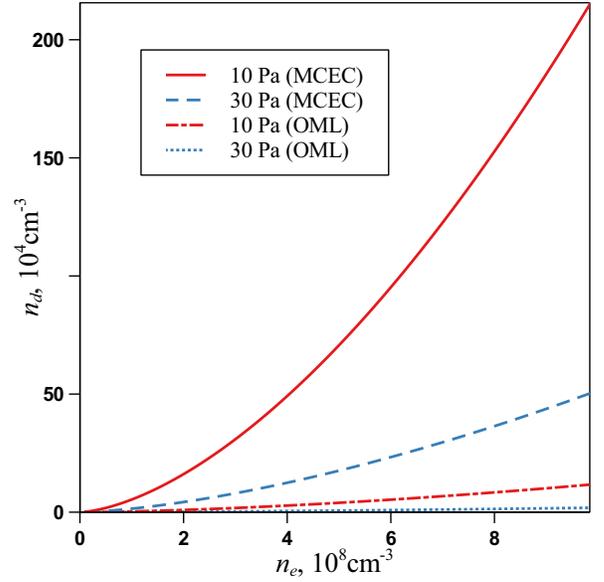}
\caption{\label{f2}Particle number density as a function of the 
electron number density at $a = 1\;\mu {\mbox{m}}$
 for argon pressure of $10$
 and $30\;{\mbox{Pa}}$
 (MCEC, solid and dashed line, respectively). Dash-dot and dot line 
indicate the OML-based calculations \cite{86} for $10$
 and $30\;{\mbox{Pa}}$, respectively. $T_e = 3.8{\kern 1pt} 
\,{\mbox{eV}}$.}
\end{figure}

\begin{figure}
\includegraphics[width=0.9\columnwidth]{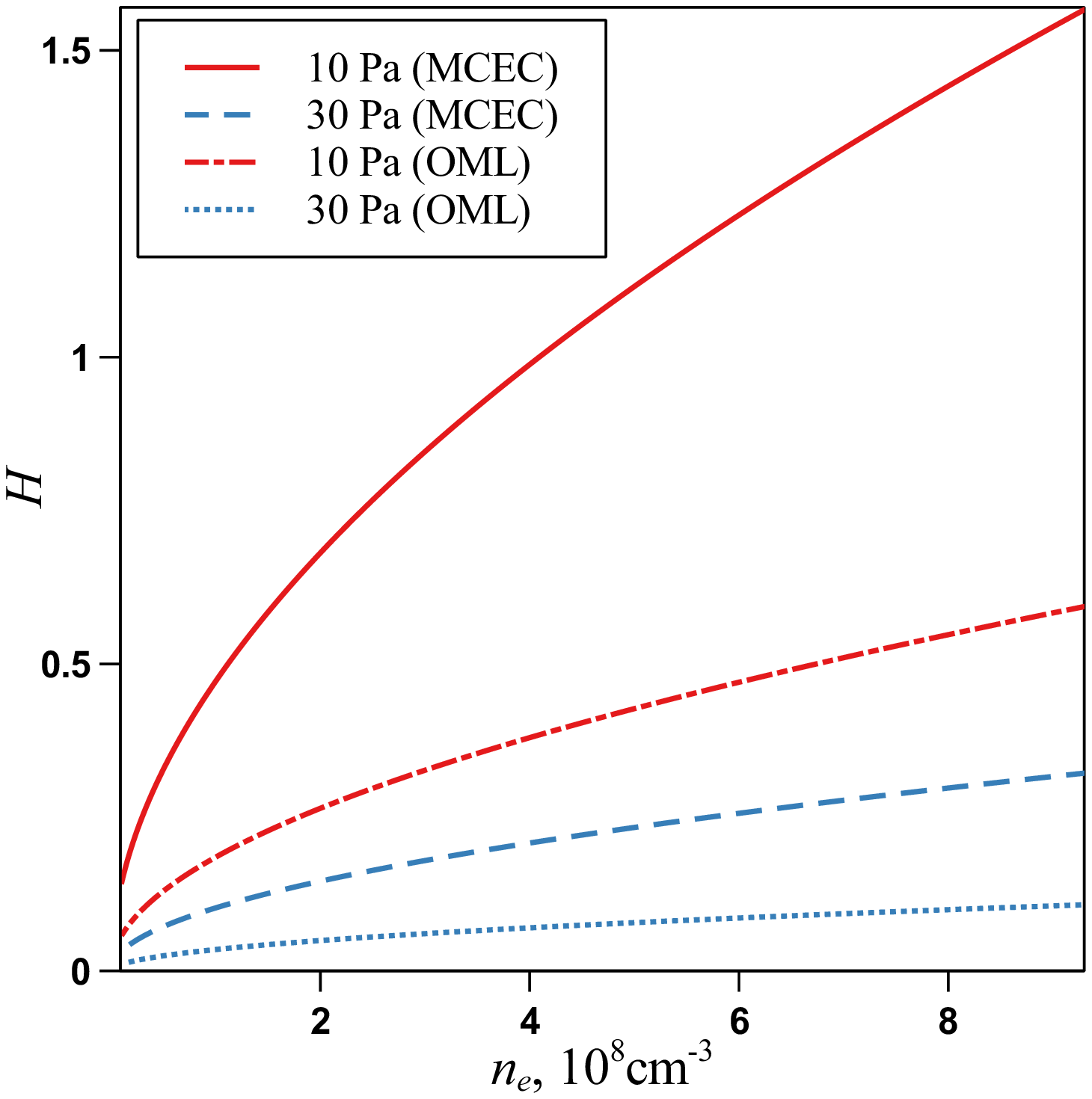}
\caption{\label{f3}Havnes parameter as a function of the electron 
number density at $a = 1\;\mu {\mbox{m}}$
 for argon pressure of $10$
 and $30\;{\mbox{Pa}}$
 (MCEC, solid and dashed line, respectively). Dash-dot and dot line 
indicate the OML-based calculations \cite{86} for $10$
 and $30\;{\mbox{Pa}}$, respectively. $T_e = 3.8{\kern 1pt} 
\,{\mbox{eV}}$.}
\end{figure}

\begin{figure}
\includegraphics[width=0.9\columnwidth]{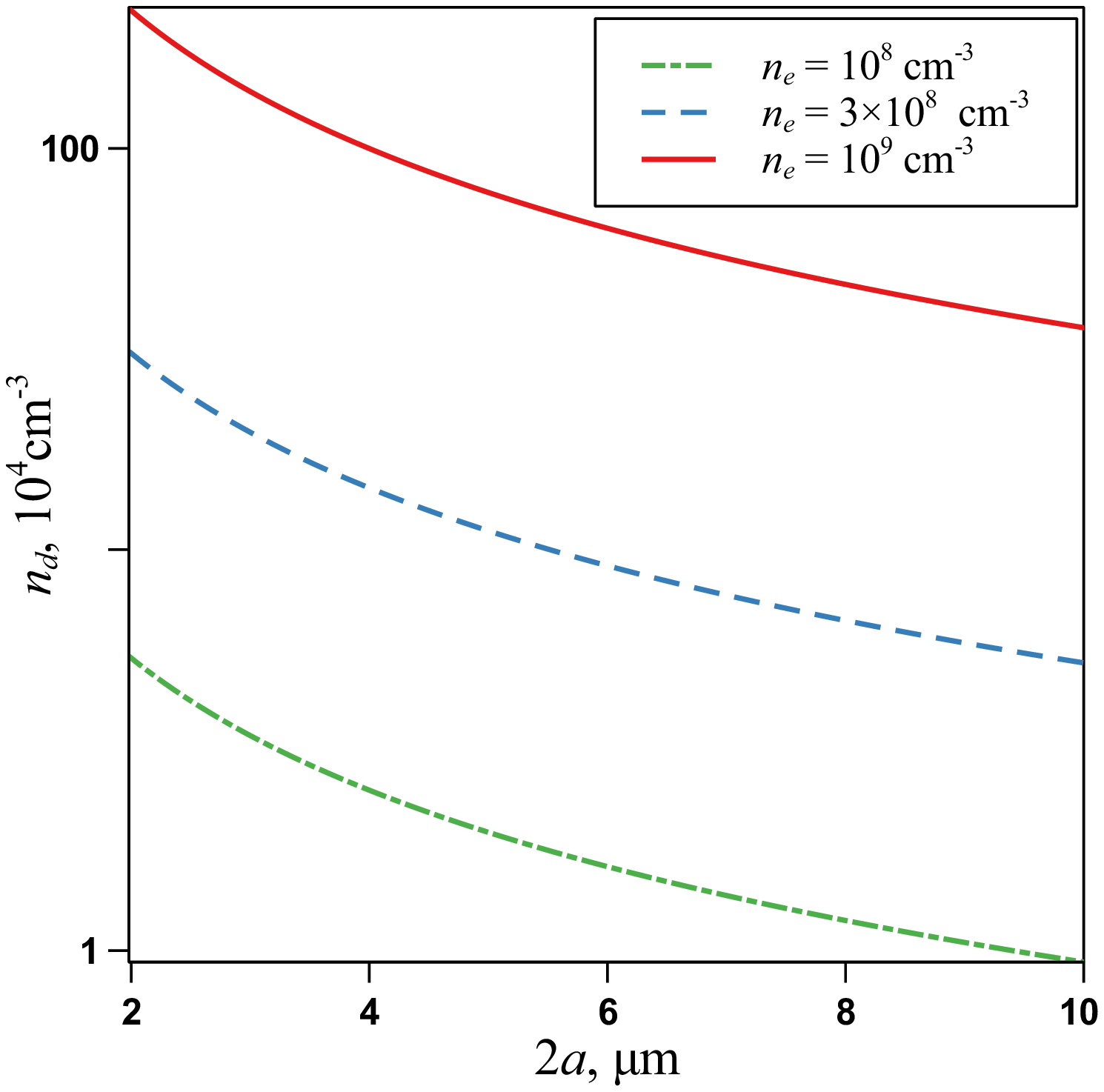}
\caption{\label{f4}Particle number density as a function of the dust 
particle diameter at $p_{\mathrm{Ar}} = 10\;{\mbox{Pa}}$
 for the electron number density of $10^8 {\kern 1pt} 
{\mbox{cm}}^{ - 3} $, $3 \times 10^8 {\kern 1pt} {\mbox{cm}}^{ - 
3} $, and $10^9 {\kern 1pt} {\mbox{cm}}^{ - 3}$ (dash-dot, dashed, 
and solid line, respectively). $T_e = 3.8{\kern 1pt} 
\,{\mbox{eV}}$.}
\end{figure}

The results of calculation based on formulas (\ref{e14}), 
(\ref{e15})--(\ref{e17}) for the discharge in argon are shown in 
Figs.~\ref{f2}--\ref{f4}. In these calculations, the ion mean free path 
is estimated as $\lambda _a = T_i /p_{\mathrm{Ar}} \sigma _{ia} $, 
where $p_{\mathrm{Ar}}$ is the argon pressure and $\sigma _{ia} 
\simeq 2 \times 10^{ - 14} \,{\mbox{cm}}^{ - 3}$ is the ion--atom 
collision cross section\cite{43}. It is seen in Fig.~\ref{f2} that the 
particle number densities corresponding to the same $n_e$ are much 
greater than in the OML approximation. Indeed, due to the particle 
charge reduction caused by the ion current enhancement, the particles 
are subject to the weaker electrostatic force. This is compensated by 
the reduction of momentum transfer cross section proportional to 
$r_d^2$ in the cell model, i.e., by the increase in $n_d $. Also, it is 
seen that $n_d$ decreases with the increase of the argon pressure, 
which flattens the dependence $n_d (n_e )$
 at high $p_{\mathrm{Ar}} $, while at low $p_{\mathrm{Ar}} $, the 
dependence $n_d (n_e )$
 is rather sharp. This effect holds in the OML approximation.

The increase of $n_d$ with the increase of $n_e$ stipulates the 
increase of the Havnes parameter (Fig.~\ref{f3}). This effect is 
especially noticeable at low argon pressure. Since the OML 
approximation leads to lower $n_d$ (cf.\ Fig.~\ref{f2}), resulting 
$H$
 is lower as well, as compared to the present approach including the 
ion current enhancement. However, note that in the OML 
approximation, $n_e$ corresponding to the experimentally measured 
$n_d$ is one or two orders of magnitude higher than that shown in 
Fig.~\ref{f3} (cf.\ \cite{69}). Eventually, in the present 
approximation, $H$
 proves to be significantly lower than that from the OML. It can be 
seen in Fig.~\ref{f3} that $H < 1$
 for $n_e$ characteristic of the available experimental data. This 
means that in many cases, one can neglect the perturbation of $n_e$ 
caused by the particles injection (this may not be true in the region 
adjacent to the void boundary because of the particle number density 
cusp \cite{69}). Thus for a quasi-homogeneous dust cloud (in the foot 
region \cite{69}), a reasonable estimate for complex plasma 
parameters can be based on the electron number density calculated for 
a discharge in a pure gas.

In the case $H \ll 1$, the particle dimensionless potential is not much 
different from its upper bound $\Phi _0$ corresponding to the limit 
$n_d \to 0$
 or $n_e \to n_i $. From (\ref{e8}), we have $\Phi _0 \simeq \theta ^{ 
- 1/3} $. Using (\ref{e9}) we obtain $\Phi _0 \simeq 0.23$
 for $a = 2 \times 10^{ - 4} \,{\mbox{cm}}$
 and $p_{\mathrm{Ar}} = 30\;{\mbox{Pa}}$. In the entire range of 
experimentally attainable argon pressures and particle diameters, $0.2 
\lesssim \Phi _0 \lesssim 0.3$, which is more than four times smaller 
than the particle potential calculated using the OML. This agrees with 
the results of recent particle charge measurements \cite{109}. 
Figure~\ref{f4} demonstrates the decreasing dependence of $n_d$ on 
the particle size at fixed $n_e $. This dependence is rather weak in 
contrast to a sharp dependence on $n_e$ illustrated again by this 
figure.

The developed IEOS's (\ref{e14}), (\ref{e15}), and (\ref{e16}) are 
based on the assumption that the Coulomb potentials of neighboring 
particles overlap; Eq.~(\ref{e11}) assumes explicitly that the 
ion--particle scattering is equivalent to the collisions of the ions 
against a hard sphere with the radius $0.45r_d $. Thus, the proposed 
model is valid if the latter does not exceed the length scale $a\tau 
\Phi$ defining the Coulomb cross section of the momentum transfer 
from an ion to the particle, i.e.,
\begin{equation}
\chi = \frac{{2.2a\tau \Phi }}{{r_d }} \ge 1. \label{e1403}
\end{equation}
As is seen in Fig.~\ref{f2}, the particle number density decreases 
with the increase of the gas pressure other parameters being fixed, 
and $r_d$ increases. Therefore, the condition (\ref{e1403}) imposes 
an upper bound on the gas pressure. If $n_e$ is decreased with the 
decreasing gas pressure then $r_d$ is increased, which implies a 
lower bound on the particle number density and the gas pressure. 
However, explicit estimates for these bounds cannot be deduced 
because the general form of the dependence $n_e (p_{\mathrm{Ar}} 
)$
 is yet unknown.

\section{\label{s5}COMPARISON WITH EXPERIMENTAL 
DATA}

Comparison of the IEOS calculation results with experimental data is 
complicated by the fact that no measurement of the electron/ion 
number density is available and that the accurate particle number 
density determination using different methods was performed only in 
\cite{43,69}. A qualitative conclusion that $n_d$ must decrease with 
the increase of $p_{\mathrm{Ar}}$ (Fig.~\ref{f2}) agrees with the 
experiment \cite{69}, according to which the dust cloud can be 
realized in two regimes. At the lower $p_{\mathrm{Ar}} $, $n_d$ 
decreases monotonically with the distance from the discharge center; 
at the higher $p_{\mathrm{Ar}} $, $n_d$ is almost constant. 
Existence of these two regimes can be accounted for by the weaker 
dependence $n_d (n_e )$
 at the higher $p_{\mathrm{Ar}} $. In addition, the spatial 
distribution of $n_e$ in a gas discharge without particles can be more 
homogeneous at the higher $p_{\mathrm{Ar}} $. Thus, the IEOS 
modification proposed in this work is capable of describing the 
dependence of the particle number density on the gas pressure.
\begin{figure}
\includegraphics[width=0.9\columnwidth]{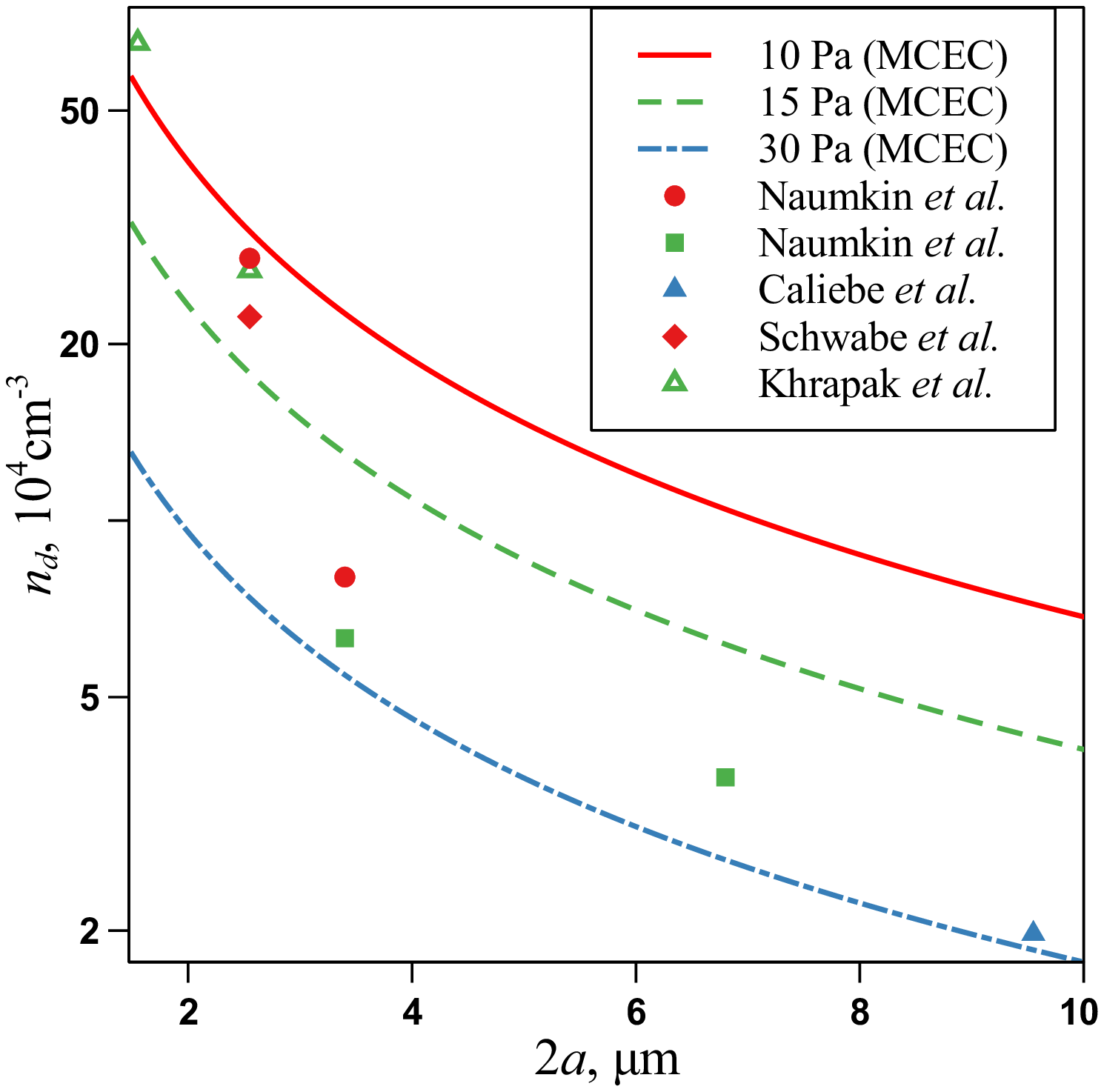}
\caption{\label{f5}Particle number density as a function of the dust 
particle diameter for the electron number density $n_e = 3.5 \times 
10^8 {\kern 1pt} {\mbox{cm}}^{ - 3}$ and different argon pressures. 
Lines indicate calculations for $p_{\mathrm{Ar}} = 
10\;{\mbox{Pa}}$, $T_e = 3.5\;{\mbox{eV}}$
 (solid line), $15\;{\mbox{Pa}}{\mbox{,}}\;\;3.8\;{\mbox{eV}}$
 (dashed line), and 
$30\;{\mbox{Pa}}{\mbox{,}}\;\;4.5\;{\mbox{eV}}$
 (dash-dot line). Dots represent experiments \cite{69} (circles for 
$p_{\mathrm{Ar}} = 10$
 and squares, for $20.5\;{\mbox{Pa}}$), \cite{11} (triangle for 
$p_{\mathrm{Ar}} = 30\;{\mbox{Pa}}$), \cite{19} (diamond for 
$p_{\mathrm{Ar}} = 10\;{\mbox{Pa}}$), and \cite{43} (open 
triangles for $p_{\mathrm{Ar}} = 15\;{\mbox{Pa}}$) (see 
Table~\ref{t1}).}
\end{figure}
\begin{table*}
\caption{\label{t1} Estimations of the Havnes parameter $H$
 Eq.~(\ref{e17}), of the parameter $\chi = 2.2a\tau \Phi /r_d $, and of 
the electron number density $n_e$ from Eqs.~(\ref{e14}), 
(\ref{e15}), and (\ref{e16}) compared to that from the OML-based 
model \cite{86} $n_e^{\mathrm{(OML)}} $, in a 
quasi-homogeneous region of the dust cloud in argon discharge based 
on the experimental data (the particle diameter $2a$, the argon 
pressure $p_{\mathrm{Ar}} $, the electron temperature $T_e $, and 
the particle number density $n_d $).}
\begin{ruledtabular}
\begin{tabular}{ccccccccc}
$2a,\;10^{ - 4} {\kern 1pt} {\mbox{cm}}$
 & $p_{\mathrm{Ar}} ,\;{\mbox{Pa}}$
 & $T_e ,\;{\mbox{eV}}$
 & $n_d ,\;10^4 {\kern 1pt} {\kern 1pt} {\mbox{cm}}^{ - 3}$ & 
Reference & $H$
 & $\chi$ & $n_e ,\;10^8 {\kern 1pt} {\kern 1pt} {\mbox{cm}}^{ - 
3}$ & $n_e^{\mathrm{(OML)}} ,\;10^8 {\kern 1pt} {\kern 1pt} 
{\mbox{cm}}^{ - 3}$ \\ \hline
1.55 & 15 & 3.8 & 65.2 & [\onlinecite{43}] & 0.723 & 1.08 & 5.60 
& 35.2 \\ 
2.55 & 15 & 3.8 & 26.7 & [\onlinecite{43}] & 0.537 & 1.17 & 4.52 
& 33.2 \\ 
9.55 & 30 & 4.5 & 1.97 & [\onlinecite{11}] & 0.113 & 1.13 & 3.64 
& 56.0 \\ 
2.55 & 10 & 3.5 & 22.3 & [\onlinecite{19}] & 0.760 & 1.17 & 2.89 
& 17.5 \\ 
2.55 & 10 & 3.5 & 28.0 & [\onlinecite{69}] & 0.818 & 1.25 & 3.27 
& 20.1 \\ 
3.4 & 11 & 3.5 & 8.01 & [\onlinecite{69}] & 0.491 & 1.04 & 1.98 & 
13.7 \\ 
3.4 & 20.5 & 3.5 & 6.30 & [\onlinecite{69}] & 0.243 & 0.84 & 2.73 
& 23.1 \\ 
6.8 & 20.5 & 3.5 & 3.65 & [\onlinecite{69}] & 0.202 & 1.14 & 3.10 
& 32.4 \\ 
\end{tabular}
\end{ruledtabular}
\end{table*}

A quantitative correspondence between the proposed IEOS and 
experiments performed with the particles of different diameters can 
be seen in Fig.~\ref{f5}. Since the electron number density at the 
point of $n_d$ measurement is unknown, we chose the common 
value of $n_e = 3.5 \times 10^8 {\kern 1pt} {\mbox{cm}}^{ - 3}$ 
most typical for the discharge in a pure gas under the same discharge 
conditions. However, the dependence $n_d (n_e )$
 is rather sharp. One should also bear in mind that each experiment is 
performed under individual argon pressure (see Table~\ref{t1}), so 
we took three average pressures, $10$, $15$, and $30\;{\mbox{Pa}}$
 to juxtapose with the experimental data, which are represented by the 
dots colored in the same way as the closest pressure in the 
calculations. In view of the foregoing, a satisfactory agreement 
between the proposed IEOS and the experimental data can be testified 
in a wide range of the particle diameter. A good reproduction of the 
trends under variation of both the gas pressure and the particle 
diameter can be seen in Fig.~\ref{f5}.

Note that only the experiments performed under microgravity 
conditions were selected for comparison with the obtained theoretical 
results in Fig.~\ref{f5}. Since the gravity adds a substantial additional 
force to those treated in the proposed model, this model cannot be 
used for the conditions of ground-based laboratory experiments. 
Thus, a correction in the theory is needed to implement it to such 
experiments.

Based on the data of experiments \cite{43,11,19,69} one can solve 
the inverse problem, i.e., calculate the electron number density at the 
point where $n_d$ was measured. The calculation results are 
summarized in Table~\ref{t1} where the corresponding Havnes 
parameter is given along with $n_e $. It can be seen that in a wide 
range of the particle diameter and number density (more than one 
order of magnitude), the resulting $n_e$ varies in a restricted range 
from $2 \times 10^8$ to $5 \times 10^8 {\kern 1pt} {\kern 1pt} 
{\mbox{cm}}^{ - 3} $. This is a consequence of the 
above-mentioned sharp dependence $n_d (n_e )$. At the same time, 
estimated $n_e$ seems to be reasonable for treated discharge 
conditions. In contrast, the electron number densities obtained from 
the OML-based IEOS \cite{86} are more than by an order of 
magnitude higher and they almost reach $10^{10} {\kern 1pt} {\kern 
1pt} {\mbox{cm}}^{ - 3} $, which seams quite unrealistic for the 
treated experimental conditions. Calculation of the parameters makes 
it possible to check the condition of MCEC validity (\ref{e1403}). It 
is seen that the condition is satisfied for almost all the experiments 
but one for $2a = 3.4\;\mu {\mbox{m}}$
 and $p_{\mathrm{Ar}} = 20.5\;{\mbox{Pa}}$. However for all 
experiments, $\chi \simeq 1$. This makes the condition (\ref{e1403}) 
compatible with $1.5a\tau \Phi /r_d = 0.68\chi \lesssim 1$, which is 
necessary to reduce (\ref{e6}) to (\ref{e7}).

Note that although $H$
 is noticeably higher for small particles, still $H < 1$, which is 
indicative of a moderate (or small) effect of the particles on the 
electron number density in argon discharge plasma.
\begin{figure}
\includegraphics[width=0.9\columnwidth]{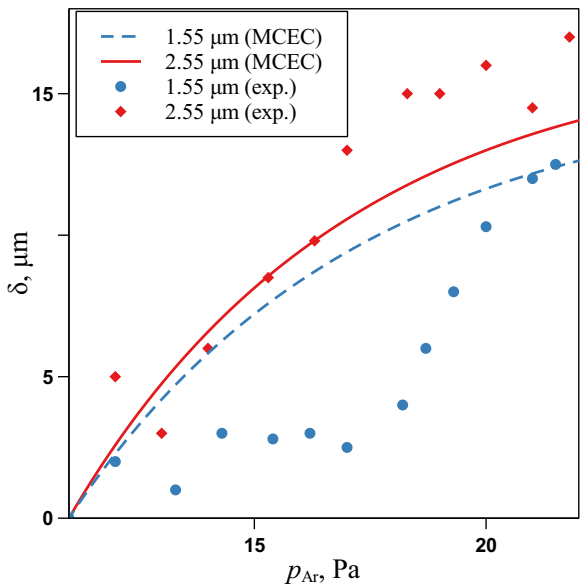}
\caption{\label{f6}Difference $\delta = n_d^{ - 1/3} - n_d^{ - 1/3} 
\left| {_{p_{\mathrm{Ar}} = 11{\kern 1pt} \mathrm{Pa}} } \right.$
 as a function of the argon pressure for $2a = 1.55\;\mu {\mbox{m}}$
 (dashed line and circles) and $2.55\;\mu {\mbox{m}}$
 (solid line and diamonds). Lines represent calculations and dots, the 
experiment\cite{43}.}
\end{figure}

In the discussion above, we considered the dependence of $n_d$ on 
$p_{\mathrm{Ar}}$ at fixed $n_e $, i.e., solely the dependence 
$\lambda _a (p_{\mathrm{Ar}} )$
 is taken into account. Instead, in a real system, $n_e$ depends on 
$p_{\mathrm{Ar}}$ as well. In the center of pure argon discharge, 
this dependence was approximated by the linear function\cite{43}
\begin{equation}
n_e = (1.20{\mbox{ }} + {\mbox{ }}0.11p_{\mathrm{Ar}} ) \times 
10^8 {\kern 1pt} {\kern 1pt} {\mbox{cm}}^{ - 3} , \label{e18}
\end{equation}
where $p_{\mathrm{Ar}}$ is in Pa. We will neglect the change of 
$n_e$ upon injection of the particles. Since $n_d$ decreases with the 
increase of $p_{\mathrm{Ar}}$ at fixed $n_e$ and increases with the 
increase of $n_e$ at fixed $p_{\mathrm{Ar}} $, the net dependence 
of $n_d$ on $p_{\mathrm{Ar}}$ is not clear if $n_e$ is related to 
$p_{\mathrm{Ar}} $. In so doing, one should bear in mind that the 
particle number density was measured in \cite{43} outside the 
discharge central region so that $n_e$ (\ref{e18}) does not coincide 
with a true electron number density at the measurement point. Hence, 
if we calculated $n_d$ with (\ref{e18}), the error could be too great. 
However, one can assume that the variation rate $dn_e 
/dp_{\mathrm{Ar}}$ is weakly dependent on the coordinate of 
measurement point. Then it is reasonable to calculate the difference 
$\delta$ between the interparticle distance $n_d^{ - 1/3} 
(p_{\mathrm{Ar}} )$
 and this quantity at some fixed pressure ($11\;{\mbox{Pa}}$). We 
used (\ref{e18}) to calculate such difference. Figure~\ref{f6} shows 
the comparison of calculation results for two particle diameters with 
the measured dependences of the interparticle distance on the 
pressure. Note that the measurement of $n_d^{ - 1/3} 
(p_{\mathrm{Ar}} )$
 \cite{43} was dynamic rather than static so that the effect of the rate 
of $p_{\mathrm{Ar}}$ variation could be nonzero. Apparently, the 
latter is responsible for a kinky arrangement of the experimental dots. 
To avoid the mess, we reproduce a single branch of the hysteresis 
corresponding to the maximum $p_{\mathrm{Ar}}$ attained for each 
particle diameter. Obviously, dynamic effects cannot be included in 
the proposed IEOS. As is seen in Fig.~\ref{f6}, the decreasing 
dependence $n_d (p_{\mathrm{Ar}} )$
 dominates the increasing dependence $n_d (n_e )$
 so that the overall effect is the decrease of $n_d$ (increase of $n_d^{ 
- 1/3} $) with the increase of $p_{\mathrm{Ar}} $. This is in a 
qualitative agreement with the experiment. Another qualitative 
correspondence is the faster increase of $n_d^{ - 1/3} 
(p_{\mathrm{Ar}} )$
 with the increase of $p_{\mathrm{Ar}}$ for the larger particles. 
Note a satisfactory quantitative agreement between the calculated and 
measured $\delta$ at the highest pressure. Furthermore, it is worth 
mentioning that the calculation formulas (\ref{e14}), (\ref{e15}), and 
(\ref{e16}) used for Figs.~\ref{f5} and \ref{f6} are free from fitting 
parameters. Therefore, they can be used for predictive calculations in 
planning the future experiments.

\section{\label{s6}SPALLATION THRESHOLD FOR THE 
DUST CLOUD}

Due to the strong Coulomb coupling, the dense cloud of dust particles 
contained in the electrostatic trap form an analog of condensed 
matter. Under certain conditions, spallation of such liquid or solid can 
occur. For example, we can consider spallation caused by the 
presence of a single probe particle of the radius $a_p \ne a$. It was 
demonstrated \cite{86} that the sum of the ion drag force and the 
electrostatic force, ${\bf{F}}_{ip}$ and ${\bf{F}}_{ep} $, 
respectively, acting on the probe particle that we term the driving 
force ${\bf{F}}_{\mathrm{drv}}$ does not vanish (${\bf{F}}_{ip} 
$, ${\bf{F}}_{ep} $, and ${\bf{F}}_{\mathrm{drv}}$ are parallel to 
the electric field strength ${\bf{E}}$). This is a result of the 
dependence of the ion mean free path on the particle number density. 
The force ${\bf{F}}_{\mathrm{drv}}$ drives the probe to the 
discharge center if $a_p < a$
 and in the opposite direction otherwise. For the following, we will 
define the direction of the coordinate axis $X$
 apart from the void center as a positive direction (this axis is parallel 
to ${\bf{E}}$) and treat the projection of the forces on $X$. If 
$F_{\mathrm{drv}}$ is sufficiently weak, the probe would diffuse 
through the cloud. In the case of a dust crystal, the diffusion would 
occur due to the local plastic deformations of a crystal. If 
$F_{\mathrm{drv}}$ exceeds some threshold, the probe would 
displace the dust particles from its rectilinear path. Thus, the dust 
particle displacement from the cylinder of the radius $R_p $, where 
$R_p$ is the radius of the probe Wigner--Seitz cell, should be 
considered. Then at the spallation threshold, the work of the driving 
force along the unit probe path $\left| {F_{\mathrm{drv}} } \right|$
 is equal to the work $\pi pR_p^2$ against the pressure $p$
 of the dust particles subsystem. Since the interparticle interaction is 
purely repulsive, $p$
 is always positive. The moving probe can thus make a space free 
from the dust particles or a lane. Apparently, this effect is similar to 
the spallation of condensed matter (e.g., upon application of a 
negative pressure). The minimum driving force, at which this effect 
can emerge, is defined by the threshold condition $Q = 1$, where
\begin{equation}
Q = \frac{{\left| {F_{\mathrm{drv}} } \right|}}{{\pi pR_p^2 }}. 
\label{e19}
\end{equation}

One can estimate the spallation criterion $Q$
 by calculation of $F_{\mathrm{drv}}$ in the same way as in 
\cite{86}. By definition, ${\bf{F}}_{\mathrm{drv}} = 
{\bf{F}}_{ep} + {\bf{F}}_{ip} $. Here, $F_{ep} = - a_p T_e \Phi _p 
E/e$, where $\Phi _p$ is the probe potential defined by the probe 
charge equation. The latter has the form [cf.\ (\ref{e8}) and 
(\ref{e9})]
\begin{equation}
\theta \Phi _p^3 \frac{{a_p }}{a}\exp \left( {\Phi _p } \right) = 1 - 
\gamma (\Phi ). \label{e20}
\end{equation}
The ion drag force is then $F_{ip} = (\pi /2)R_p^2 n_i \lambda _p 
eE$, where $R_p$ is defined by the relation \cite{22}
\begin{equation}
R_p^2 = \frac{{Z_p e^2 }}{{(8\pi p)^{1/2} }} = \left( {\frac{{\rho 
\lambda _a }}{S}} \right)^2 , \label{e21}
\end{equation}
and $S = \left( {a\Phi /a_p \Phi _p } \right)^{1/2} $. Here, we took 
into account that $\Phi _p = Z_p e^2 /a_p T_e $. In contrast to 
Eq.~(18) in \cite{86}, Eq.~(\ref{e21}) includes the ratio $\Phi _p 
/\Phi $. The local ion mean free path in the vicinity of a probe is 
defined by the approximation similar to (\ref{e10}) (cf.\ \cite{86})
\begin{equation}
\lambda _p = \lambda _a \left( {1 + \frac{{3\lambda _a }}{{8R_p 
}}} \right)^{ - 1} = \lambda _a \left( {1 + \frac{{3S}}{{8\rho }}} 
\right)^{ - 1} . \label{e22}
\end{equation}
Thus, we obtain
\begin{equation}
\kappa \equiv \frac{{F_{\mathrm{drv}} }}{{F_{ep} }} = \frac{{S - 
1}}{{S + 8\rho /3}}, \label{e23}
\end{equation}
and the driving force is
\begin{equation}
F_{\mathrm{drv}} = \frac{{\kappa a_p \Phi _p T_e^2 }}{{Le^2 }}, 
\label{e24}
\end{equation}
where $L = \left| {\bm{\nabla}\ln n_e } \right|^{ - 1}$ is the length 
scale of the electron number density variation. Based on (\ref{e19}), 
(\ref{e23}), (\ref{e24}), and the estimation for the particle pressure 
\cite{22} $p = Z^2 e^2 /8\pi r_d^4$ we derive eventually the 
spallation criterion
\begin{equation}
Q = \frac{{8\left| \kappa \right|r_d^2 }}{{a\Phi L}}, \label{e25}
\end{equation}
which can be calculated for a stationary dust cloud on the basis of the 
IEOS's (\ref{e14}), (\ref{e15}), and (\ref{e16}).
\begin{figure}
\includegraphics[width=0.9\columnwidth]{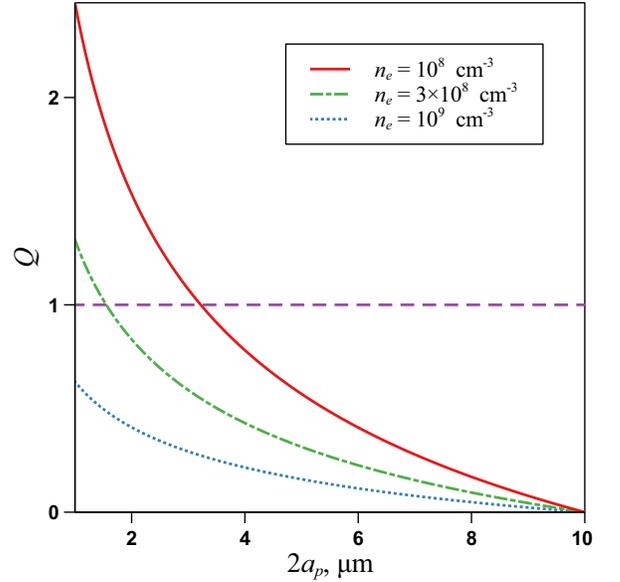}
\caption{\label{f7}Spallation criterion (\ref{e25}) as a function of the 
probe particle diameter for the electron number density of $10^8 
{\kern 1pt} {\mbox{cm}}^{ - 3} $, $3 \times 10^8 {\kern 1pt} 
{\mbox{cm}}^{ - 3} $, and $10^9 {\kern 1pt} {\mbox{cm}}^{ - 3}$ 
(sold, dash-dot, and dot line, respectively), $2a = 10\;\mu 
{\mbox{m}}$, and $p_{\mathrm{Ar}} = 30\;{\mbox{Pa}}$. Dashed 
line indicates the spallation threshold. $L = 5\;{\mbox{cm}}$
 and $T_e = 3.8{\kern 1pt} \,{\mbox{eV}}$.}
\end{figure}
\begin{figure}
\includegraphics[width=0.9\columnwidth]{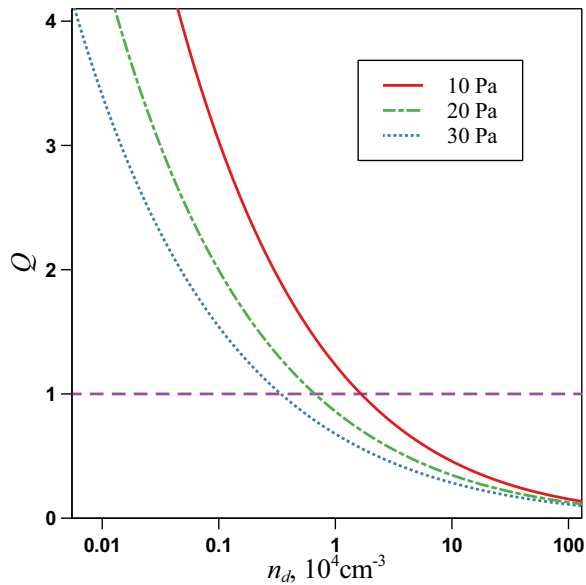}
\caption{\label{f8}Spallation criterion (\ref{e25}) as a function of the 
dust particle number density for $p_{\mathrm{Ar}} = 10$, $20$, and 
$30\;{\mbox{Pa}}$
 (sold, dash-dot, and dot line, respectively), $2a = 10\;\mu 
{\mbox{m}}$, and $2a_p = 3\;\mu {\mbox{m}}$. Dashed line 
indicates the spallation threshold. $L = 5\;{\mbox{cm}}$
 and $T_e = 3.8{\kern 1pt} \,{\mbox{eV}}$.}
\end{figure}

If the probe radius is close to that of the dust particles, $\left| {a - a_p 
} \right|/a \ll 1$
 then one can use (\ref{e23}) to write approximately
\begin{equation}
\kappa \simeq \frac{{\lambda _a }}{{8r_d + 3\lambda _a }}\left( {1 - 
\frac{{a_p }}{a}} \right). \label{e26}
\end{equation}
From (\ref{e25}) and (\ref{e26}) for $\left| {a - a_p } \right|/a \sim 1$
 and $\rho > 3/8$, one can obtain a crude estimate $Q \simeq r_d 
\lambda _a /a\Phi L$. This means that spallation would be impossible 
for a dense system (small $r_d $) of large particles at high argon 
pressure. Since $Q \propto L^{ - 1} \propto E$, the ambipolar electric 
field must be sufficiently strong. In addition, the increase of 
$p_{\mathrm{Ar}}$ increases $r_d$ but decreases $\lambda _a$ and 
$\Phi $. As a result, $Q$
 is almost independent of the argon pressure at fixed $n_e $.

\begin{figure}
\includegraphics[width=0.9\columnwidth]{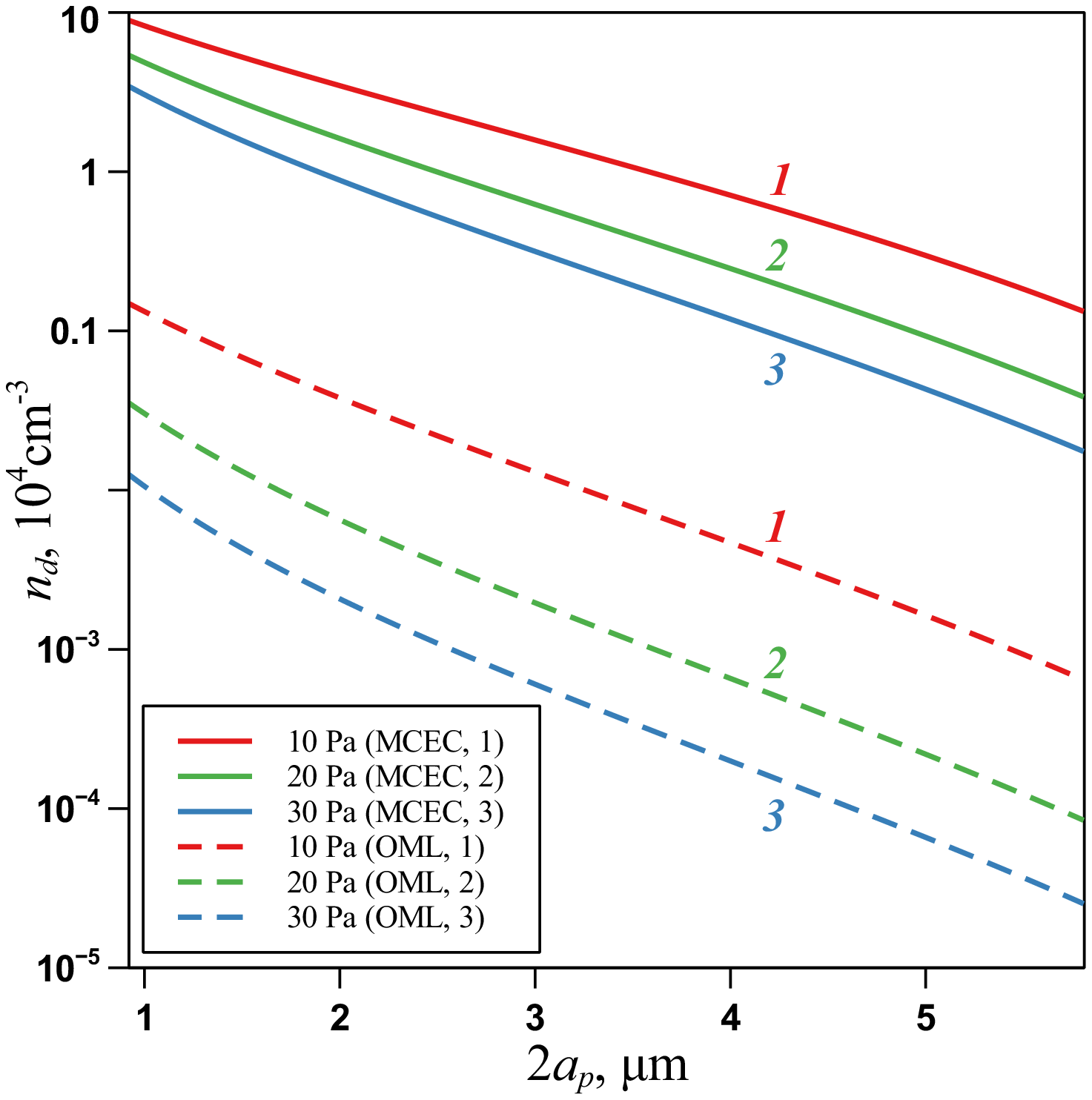}
\caption{\label{f9}Particle number density along the line $Q = 1$
 (spallation threshold) as a function of the probe particle diameter for 
$2a = 9.2\;\mu {\mbox{m}}$
 and the argon pressure of ({\it 1}) $10$, ({\it 2}) $20$, and ({\it 3}) 
$30\;{\mbox{Pa}}$
 (MCEC). Dashed lines indicate similar results from the OML-based 
IEOS \cite{86} for the argon pressure of ({\it 1}) $10$, ({\it 2}) 
$20$, and ({\it 3}) $30\;{\mbox{Pa}}$. $L = 5\;{\mbox{cm}}$
 and $T_e = 3.8{\kern 1pt} \,{\mbox{eV}}$.}
\end{figure}

The objective of the following calculations is to clarify the 
conditions, under which spallation is favored. In so doing, we will 
confine ourselves to the case $a > a_p$ and assume a 
quasi-homogeneous dust cloud in the argon discharge. Then the 
estimation of the length scale $L \sim 5\;{\mbox{cm}}$
 in (\ref{e25}) can be of the same order as typical discharge 
dimensions. Figure~\ref{f7} illustrates the spallation accessibility for 
different probe diameters provided that the dust particle diameter and 
the argon pressure are fixed. As is seen, spallation is impossible as 
$a_p \to a$; this follows straightforwardly from (\ref{e25}) and 
(\ref{e26}). Spallation is hindered for high electron number density 
as well because of the associated increase of $n_d $. It follows from 
Fig.~\ref{f7} that spallation is possible for a probe of the diameter 
about $4\;\mu {\mbox{m}}$
 at the typical level of $n_e \sim 10^8 \,{\mbox{cm}}^{ - 3} $. 
Figure~\ref{f8} shows that the increase of the argon pressure other 
parameters being fixed increases $Q$. The particle number density, at 
which spallation is possible, ranges between $10^3$ and $10^4 
\,{\mbox{cm}}^{ - 3} $.

The set of Eqs.~ (\ref{e14}), (\ref{e15}), and (\ref{e16}) combined 
with the threshold condition $Q = 1$
 define uniquely the parameters of complex plasma such as the 
threshold particle number density for given probe diameter 
(Fig.~\ref{f9}). This figure demonstrates that the decrease in argon 
pressure favors spallation, i.e., it shifts the threshold line toward the 
region of the denser and more strongly coupled system. It is seen that 
for $2a_p = 3.4\;\mu {\mbox{m}}$, $2a = 9.2\;\mu {\mbox{m}}$, 
and $p_{\mathrm{Ar}} = 30\;{\mbox{Pa}}$, the threshold particle 
number density $10^3 \lesssim n_d \lesssim 10^4 \,{\mbox{cm}}^{ - 
3} $. This agrees with the result of experiment \cite{68}, in which a 
beam of smaller particles penetrated a quasi-homogeneous stationary 
cloud of the larger dust particles thus forming lanes. The particle 
diameters and argon pressure are specified above. Judging from 
Fig.~1 of \cite{68}, the experimental particle number density can be 
in the same range from ca.\ $10^3$ to $10^4 \,{\mbox{cm}}^{ - 3} $. 
For comparison, Fig.~\ref{f9} also shows the line $Q = 1$, where the 
quantities included in (\ref{e25}) are calculated using the OML-based 
IEOS \cite{86}. It is seen that the threshold particle number densities 
are up to three orders of magnitude lower than those calculated in this 
work. Thus, consideration of the ion--atom collisions is 
fundamentally important for the treatment of lane formation.

Note that the spallation criterion for a beam of particles can be 
different from that for a single probe particle. In addition, it is a 
matter of discussion whether the lane formation observed in \cite{68} 
can be treated as the spallation.

\section{\label{s7} CONCLUSION}

In this study, we propose a modification of the IEOS that includes the 
effect of the ion--atom collisions in the vicinity of dust particles. 
Toward this end, we estimated the screening length for the ``dense'' 
dust cloud in the framework of the Wigner--Seitz cell model. This 
screening length proved to be typically larger than the ion Debye 
screening length. Fortunately, it cancels in the expression for ion 
current to the particle (\ref{e7}) and, correspondingly, in the 
equations for the particle charge (\ref{e8}), (\ref{e9}). Inclusion of 
the ion--atom collisions leads to more than an order of magnitude 
increase in the estimated ion current (this effect is proportional to 
$a/\lambda _a $), which implies the decrease of the particle charge 
$Z$. It was demonstrated that the necessary condition to treat 
collisionless plasma is $\lambda _a /r_d > 10$, which demonstrates 
the importance of included effect.

The IEOS's for a dense dust cloud in the low-pressure gas discharge 
are based on the particle charge equation, the quasineutrality 
equation, and the balance equation for the electrostatic force and the 
ion drag force acting on a particle. The latter force takes into account 
both the effects of the ion--atom and that of the ion--particle 
collisions. It follows from obtained IEOS's that the particle number 
density decreases with the increase of the particle diameter and the 
gas pressure and it increases rather sharply with the increase of the 
electron number density. Since in a real discharge, the latter is, in its 
turn, dependent on the gas pressure, the performed calculations took 
into account this dependence. Comparison between the theory and 
available experimental data concerning the particle number densities 
is indicative of a satisfactory quantitative agreement in a wide range 
of variation of complex plasma parameters. In particular, calculations 
demonstrate the net effect of the decrease of $n_d$ with the increase 
of $p_{\mathrm{Ar}} $. Note that used IEOS's includes no fitting 
parameters.

The following shortcomings of the proposed theory should be noted. 
The inapplicability of this theory for the interpretation of 
ground-based experiments has already been noted in Sec.~\ref{s5}. 
Next, the implementation of IEOS's implies that one of the complex 
plasma parameters is known. Calculation of all plasma parameters 
would require incorporation of IEOS's with the equations for the 
ionization kinetics. Then, the proposed theory is local, which implies 
that all quantities are at least continuous. This is not true in the 
vicinity of the void boundary (in the cusp region \cite{69}), where 
$n_d$ changes abruptly. In this region, the theory is invalid.

Obtained IEOS's proved to have sufficient accuracy to estimate the 
threshold of the lane formation, which is sensitive to the plasma 
parameters. We assume that emergence of the lanes upon injection of 
small dust particles (probes) into a cloud of large particles is a 
manifestation of spallation of the plasma crystal caused by the probes. 
In contrast to the lane formation in colloidal mixtures, this could 
rather be similar to spinodal decay than to a nonequilibrium phase 
transition. The probe particle in a dust cloud that is of the size 
different from that of the cloud particles is subject to the driving 
force, which is a result of the dependence of the local ion mean free 
path and, correspondingly, of the ion driving force on the particle 
size. The moving probe particle can form a cylindrical cavity if the 
work of driving force is greater than the work against the positive 
pressure of the cloud particles. This enables one to obtain the 
spallation criterion that can be calculated on the basis of IEOS's. The 
calculations show that the lane formation is possible provided that the 
size difference between the probe and the cloud particles is 
sufficiently large, and if the electron and particle number density is 
sufficiently low. The lane formation onset criterion increases, i.e., the 
threshold decreases, with the decreasing gas pressure. We 
demonstrate that under the conditions of experiment \cite{68}, the 
threshold number density of the cloud particles must be in the interval 
$10^3 \lesssim n_d \lesssim 10^4 \,{\mbox{cm}}^{ - 3} $, which 
agrees with experimental data. Apparently, this interval can be typical 
for similar experiments. Since no experimental information on the 
lane formation threshold is available, conducting new experiments, in 
which this threshold can be measured and its dependence on plasma 
parameters can be determined, is an urgent task in this field. The 
theory proposed in this work treats solely individual probes and does 
not take into account the collective motion of such particles. The open 
issues mentioned above will be addressed in the future.

\begin{acknowledgments}
This work was supported by Presidium RAS program No.\ 13 
``Condensed Matter and Plasma at High Energy Densities''.
\end{acknowledgments}

\providecommand{\noopsort}[1]{}\providecommand{\singleletter}[1]{#1}%


\begin{thebibliography}{38}%
\makeatletter
\providecommand \@ifxundefined [1]{%
 \@ifx{#1\undefined}
}%
\providecommand \@ifnum [1]{%
 \ifnum #1\expandafter \@firstoftwo
 \else \expandafter \@secondoftwo
 \fi
}%
\providecommand \@ifx [1]{%
 \ifx #1\expandafter \@firstoftwo
 \else \expandafter \@secondoftwo
 \fi
}%
\providecommand \natexlab [1]{#1}%
\providecommand \enquote  [1]{``#1''}%
\providecommand \bibnamefont  [1]{#1}%
\providecommand \bibfnamefont [1]{#1}%
\providecommand \citenamefont [1]{#1}%
\providecommand \href@noop [0]{\@secondoftwo}%
\providecommand \href [0]{\begingroup \@sanitize@url \@href}%
\providecommand \@href[1]{\@@startlink{#1}\@@href}%
\providecommand \@@href[1]{\endgroup#1\@@endlink}%
\providecommand \@sanitize@url [0]{\catcode `\\12\catcode `\$12\catcode
  `\&12\catcode `\#12\catcode `\^12\catcode `\_12\catcode `\%12\relax}%
\providecommand \@@startlink[1]{}%
\providecommand \@@endlink[0]{}%
\providecommand \url  [0]{\begingroup\@sanitize@url \@url }%
\providecommand \@url [1]{\endgroup\@href {#1}{\urlprefix }}%
\providecommand \urlprefix  [0]{URL }%
\providecommand \Eprint [0]{\href }%
\providecommand \doibase [0]{http://dx.doi.org/}%
\providecommand \selectlanguage [0]{\@gobble}%
\providecommand \bibinfo  [0]{\@secondoftwo}%
\providecommand \bibfield  [0]{\@secondoftwo}%
\providecommand \translation [1]{[#1]}%
\providecommand \BibitemOpen [0]{}%
\providecommand \bibitemStop [0]{}%
\providecommand \bibitemNoStop [0]{.\EOS\space}%
\providecommand \EOS [0]{\spacefactor3000\relax}%
\providecommand \BibitemShut  [1]{\csname bibitem#1\endcsname}%
\let\auto@bib@innerbib\@empty
\bibitem [{\citenamefont {Fortov}\ and\ \citenamefont {Morfill}(2010)}]{1}%
  \BibitemOpen
  \bibinfo {editor} {\bibfnamefont {V.~E.}\ \bibnamefont {Fortov}}\ and\
  \bibinfo {editor} {\bibfnamefont {G.~E.}\ \bibnamefont {Morfill}},\ eds.,\
  \href@noop {} {\emph {\bibinfo {title} {Complex and Dusty Plasmas: From
  Laboratory to Space}}},\ Series in Plasma Physics\ (\bibinfo  {publisher}
  {CRC Press},\ \bibinfo {address} {Boca Raton, FL},\ \bibinfo {year}
  {2010})\BibitemShut {NoStop}%
\bibitem [{\citenamefont {Chu}\ and\ \citenamefont {Lin}(1994)}]{2}%
  \BibitemOpen
  \bibfield  {author} {\bibinfo {author} {\bibfnamefont {J.~H.}\ \bibnamefont
  {Chu}}\ and\ \bibinfo {author} {\bibfnamefont {I.}~\bibnamefont {Lin}},\
  }\href {\doibase 10.1103/PhysRevLett.72.4009} {\bibfield  {journal} {\bibinfo
   {journal} {Phys.\ Rev.\ Lett.}\ }\textbf {\bibinfo {volume} {72}},\ \bibinfo
  {pages} {4009} (\bibinfo {year} {1994})}\BibitemShut {NoStop}%
\bibitem [{\citenamefont {Thomas}\ \emph {et~al.}(1994)\citenamefont {Thomas},
  \citenamefont {Morfill}, \citenamefont {Demmel}, \citenamefont {Goree},
  \citenamefont {Feuerbacher},\ and\ \citenamefont {M{\"{o}}hlmann}}]{3}%
  \BibitemOpen
  \bibfield  {author} {\bibinfo {author} {\bibfnamefont {H.}~\bibnamefont
  {Thomas}}, \bibinfo {author} {\bibfnamefont {G.~E.}\ \bibnamefont {Morfill}},
  \bibinfo {author} {\bibfnamefont {V.}~\bibnamefont {Demmel}}, \bibinfo
  {author} {\bibfnamefont {J.}~\bibnamefont {Goree}}, \bibinfo {author}
  {\bibfnamefont {B.}~\bibnamefont {Feuerbacher}}, \ and\ \bibinfo {author}
  {\bibfnamefont {D.}~\bibnamefont {M{\"{o}}hlmann}},\ }\href {\doibase
  10.1103/PhysRevLett.73.652} {\bibfield  {journal} {\bibinfo  {journal}
  {Phys.\ Rev.\ Lett.}\ }\textbf {\bibinfo {volume} {73}},\ \bibinfo {pages}
  {652} (\bibinfo {year} {1994})}\BibitemShut {NoStop}%
\bibitem [{\citenamefont {Vladimirov}\ \emph {et~al.}(2005)\citenamefont
  {Vladimirov}, \citenamefont {Ostrikov},\ and\ \citenamefont {Samarian}}]{5}%
  \BibitemOpen
  \bibfield  {author} {\bibinfo {author} {\bibfnamefont {S.~V.}\ \bibnamefont
  {Vladimirov}}, \bibinfo {author} {\bibfnamefont {K.}~\bibnamefont
  {Ostrikov}}, \ and\ \bibinfo {author} {\bibfnamefont {A.~A.}\ \bibnamefont
  {Samarian}},\ }\href@noop {} {\emph {\bibinfo {title} {Physics and
  Applications of Complex Plasmas}}}\ (\bibinfo  {publisher} {Imperial
  College},\ \bibinfo {address} {London},\ \bibinfo {year} {2005})\BibitemShut
  {NoStop}%
\bibitem [{\citenamefont {Fortov}\ \emph {et~al.}(2005)\citenamefont {Fortov},
  \citenamefont {Ivlev}, \citenamefont {Khrapak}, \citenamefont {Khrapak},\
  and\ \citenamefont {Morfill}}]{6}%
  \BibitemOpen
  \bibfield  {author} {\bibinfo {author} {\bibfnamefont {V.}~\bibnamefont
  {Fortov}}, \bibinfo {author} {\bibfnamefont {A.}~\bibnamefont {Ivlev}},
  \bibinfo {author} {\bibfnamefont {S.}~\bibnamefont {Khrapak}}, \bibinfo
  {author} {\bibfnamefont {A.}~\bibnamefont {Khrapak}}, \ and\ \bibinfo
  {author} {\bibfnamefont {G.}~\bibnamefont {Morfill}},\ }\href {\doibase
  10.1016/j.physrep.2005.08.007} {\bibfield  {journal} {\bibinfo  {journal}
  {Phys.\ Rep.}\ }\textbf {\bibinfo {volume} {421}},\ \bibinfo {pages} {1}
  (\bibinfo {year} {2005})}\BibitemShut {NoStop}%
\bibitem [{\citenamefont {Bonitz}\ \emph {et~al.}(2010)\citenamefont {Bonitz},
  \citenamefont {Henning},\ and\ \citenamefont {Block}}]{9}%
  \BibitemOpen
  \bibfield  {author} {\bibinfo {author} {\bibfnamefont {M.}~\bibnamefont
  {Bonitz}}, \bibinfo {author} {\bibfnamefont {C.}~\bibnamefont {Henning}}, \
  and\ \bibinfo {author} {\bibfnamefont {D.}~\bibnamefont {Block}},\ }\href
  {\doibase 10.1088/0034-4885/73/6/066501} {\bibfield  {journal} {\bibinfo
  {journal} {Rep.\ Prog.\ Phys.}\ }\textbf {\bibinfo {volume} {73}},\ \bibinfo
  {pages} {066501} (\bibinfo {year} {2010})}\BibitemShut {NoStop}%
\bibitem [{\citenamefont {Morfill}\ \emph {et~al.}(2006)\citenamefont
  {Morfill}, \citenamefont {Konopka}, \citenamefont {Kretschmer}, \citenamefont
  {Rubin-Zuzic}, \citenamefont {Thomas}, \citenamefont {Zhdanov},\ and\
  \citenamefont {Tsytovich}}]{10}%
  \BibitemOpen
  \bibfield  {author} {\bibinfo {author} {\bibfnamefont {G.~E.}\ \bibnamefont
  {Morfill}}, \bibinfo {author} {\bibfnamefont {U.}~\bibnamefont {Konopka}},
  \bibinfo {author} {\bibfnamefont {M.}~\bibnamefont {Kretschmer}}, \bibinfo
  {author} {\bibfnamefont {M.}~\bibnamefont {Rubin-Zuzic}}, \bibinfo {author}
  {\bibfnamefont {H.~M.}\ \bibnamefont {Thomas}}, \bibinfo {author}
  {\bibfnamefont {S.~K.}\ \bibnamefont {Zhdanov}}, \ and\ \bibinfo {author}
  {\bibfnamefont {V.}~\bibnamefont {Tsytovich}},\ }\href {\doibase
  10.1088/1367-2630/8/1/007} {\bibfield  {journal} {\bibinfo  {journal} {New
  J.\ Phys.}\ }\textbf {\bibinfo {volume} {8}},\ \bibinfo {pages} {7} (\bibinfo
  {year} {2006})}\BibitemShut {NoStop}%
\bibitem [{\citenamefont {Schwabe}\ \emph {et~al.}(2008)\citenamefont
  {Schwabe}, \citenamefont {Zhdanov}, \citenamefont {Thomas}, \citenamefont
  {Ivlev}, \citenamefont {Rubin-Zuzic}, \citenamefont {Morfill}, \citenamefont
  {Molotkov}, \citenamefont {Lipaev}, \citenamefont {Fortov},\ and\
  \citenamefont {Reiter}}]{15}%
  \BibitemOpen
  \bibfield  {author} {\bibinfo {author} {\bibfnamefont {M.}~\bibnamefont
  {Schwabe}}, \bibinfo {author} {\bibfnamefont {S.~K.}\ \bibnamefont
  {Zhdanov}}, \bibinfo {author} {\bibfnamefont {H.~M.}\ \bibnamefont {Thomas}},
  \bibinfo {author} {\bibfnamefont {A.~V.}\ \bibnamefont {Ivlev}}, \bibinfo
  {author} {\bibfnamefont {M.}~\bibnamefont {Rubin-Zuzic}}, \bibinfo {author}
  {\bibfnamefont {G.~E.}\ \bibnamefont {Morfill}}, \bibinfo {author}
  {\bibfnamefont {V.~I.}\ \bibnamefont {Molotkov}}, \bibinfo {author}
  {\bibfnamefont {A.~M.}\ \bibnamefont {Lipaev}}, \bibinfo {author}
  {\bibfnamefont {V.~E.}\ \bibnamefont {Fortov}}, \ and\ \bibinfo {author}
  {\bibfnamefont {T.}~\bibnamefont {Reiter}},\ }\href {\doibase
  10.1088/1367-2630/10/3/033037} {\bibfield  {journal} {\bibinfo  {journal}
  {New J.\ Phys.}\ }\textbf {\bibinfo {volume} {10}},\ \bibinfo {pages}
  {033037} (\bibinfo {year} {2008})}\BibitemShut {NoStop}%
\bibitem [{\citenamefont {Morfill}\ \emph {et~al.}(1999)\citenamefont
  {Morfill}, \citenamefont {Thomas}, \citenamefont {Konopka}, \citenamefont
  {Rothermel}, \citenamefont {Zuzic}, \citenamefont {Ivlev},\ and\
  \citenamefont {Goree}}]{16}%
  \BibitemOpen
  \bibfield  {author} {\bibinfo {author} {\bibfnamefont {G.~E.}\ \bibnamefont
  {Morfill}}, \bibinfo {author} {\bibfnamefont {H.~M.}\ \bibnamefont {Thomas}},
  \bibinfo {author} {\bibfnamefont {U.}~\bibnamefont {Konopka}}, \bibinfo
  {author} {\bibfnamefont {H.}~\bibnamefont {Rothermel}}, \bibinfo {author}
  {\bibfnamefont {M.}~\bibnamefont {Zuzic}}, \bibinfo {author} {\bibfnamefont
  {A.}~\bibnamefont {Ivlev}}, \ and\ \bibinfo {author} {\bibfnamefont
  {J.}~\bibnamefont {Goree}},\ }\href {\doibase 10.1103/PhysRevLett.83.1598}
  {\bibfield  {journal} {\bibinfo  {journal} {Phys.\ Rev.\ Lett.}\ }\textbf
  {\bibinfo {volume} {83}},\ \bibinfo {pages} {1598} (\bibinfo {year}
  {1999})}\BibitemShut {NoStop}%
\bibitem [{\citenamefont {Khrapak}\ \emph {et~al.}(2011)\citenamefont
  {Khrapak}, \citenamefont {Klumov}, \citenamefont {Huber}, \citenamefont
  {Molotkov}, \citenamefont {Lipaev}, \citenamefont {Naumkin}, \citenamefont
  {Thomas}, \citenamefont {Ivlev}, \citenamefont {Morfill}, \citenamefont
  {Petrov}, \citenamefont {Fortov}, \citenamefont {Malentschenko},\ and\
  \citenamefont {Volkov}}]{17}%
  \BibitemOpen
  \bibfield  {author} {\bibinfo {author} {\bibfnamefont {S.~A.}\ \bibnamefont
  {Khrapak}}, \bibinfo {author} {\bibfnamefont {B.~A.}\ \bibnamefont {Klumov}},
  \bibinfo {author} {\bibfnamefont {P.}~\bibnamefont {Huber}}, \bibinfo
  {author} {\bibfnamefont {V.~I.}\ \bibnamefont {Molotkov}}, \bibinfo {author}
  {\bibfnamefont {A.~M.}\ \bibnamefont {Lipaev}}, \bibinfo {author}
  {\bibfnamefont {V.~N.}\ \bibnamefont {Naumkin}}, \bibinfo {author}
  {\bibfnamefont {H.~M.}\ \bibnamefont {Thomas}}, \bibinfo {author}
  {\bibfnamefont {A.~V.}\ \bibnamefont {Ivlev}}, \bibinfo {author}
  {\bibfnamefont {G.~E.}\ \bibnamefont {Morfill}}, \bibinfo {author}
  {\bibfnamefont {O.~F.}\ \bibnamefont {Petrov}}, \bibinfo {author}
  {\bibfnamefont {V.~E.}\ \bibnamefont {Fortov}}, \bibinfo {author}
  {\bibfnamefont {{\relax Yu}.}~\bibnamefont {Malentschenko}}, \ and\ \bibinfo
  {author} {\bibfnamefont {S.}~\bibnamefont {Volkov}},\ }\href {\doibase
  10.1103/PhysRevLett.106.205001} {\bibfield  {journal} {\bibinfo  {journal}
  {Phys.\ Rev.\ Lett.}\ }\textbf {\bibinfo {volume} {106}},\ \bibinfo {pages}
  {205001} (\bibinfo {year} {2011})}\BibitemShut {NoStop}%
\bibitem [{\citenamefont {Thomas}\ \emph {et~al.}(2008)\citenamefont {Thomas},
  \citenamefont {Morfill}, \citenamefont {Fortov}, \citenamefont {Ivlev},
  \citenamefont {Molotkov}, \citenamefont {Lipaev}, \citenamefont {Hagl},
  \citenamefont {Rothermel}, \citenamefont {Khrapak}, \citenamefont
  {Suetterlin}, \citenamefont {Rubin-Zuzic}, \citenamefont {Petrov},
  \citenamefont {Tokarev},\ and\ \citenamefont {Krikalev}}]{18}%
  \BibitemOpen
  \bibfield  {author} {\bibinfo {author} {\bibfnamefont {H.~M.}\ \bibnamefont
  {Thomas}}, \bibinfo {author} {\bibfnamefont {G.~E.}\ \bibnamefont {Morfill}},
  \bibinfo {author} {\bibfnamefont {V.~E.}\ \bibnamefont {Fortov}}, \bibinfo
  {author} {\bibfnamefont {A.~V.}\ \bibnamefont {Ivlev}}, \bibinfo {author}
  {\bibfnamefont {V.~I.}\ \bibnamefont {Molotkov}}, \bibinfo {author}
  {\bibfnamefont {A.~M.}\ \bibnamefont {Lipaev}}, \bibinfo {author}
  {\bibfnamefont {T.}~\bibnamefont {Hagl}}, \bibinfo {author} {\bibfnamefont
  {H.}~\bibnamefont {Rothermel}}, \bibinfo {author} {\bibfnamefont {S.~A.}\
  \bibnamefont {Khrapak}}, \bibinfo {author} {\bibfnamefont {R.~K.}\
  \bibnamefont {Suetterlin}}, \bibinfo {author} {\bibfnamefont
  {M.}~\bibnamefont {Rubin-Zuzic}}, \bibinfo {author} {\bibfnamefont {O.~F.}\
  \bibnamefont {Petrov}}, \bibinfo {author} {\bibfnamefont {V.~I.}\
  \bibnamefont {Tokarev}}, \ and\ \bibinfo {author} {\bibfnamefont {S.~K.}\
  \bibnamefont {Krikalev}},\ }\href {\doibase 10.1088/1367-2630/10/3/033036}
  {\bibfield  {journal} {\bibinfo  {journal} {New J.\ Phys.}\ }\textbf
  {\bibinfo {volume} {10}},\ \bibinfo {pages} {033036} (\bibinfo {year}
  {2008})}\BibitemShut {NoStop}%
\bibitem [{\citenamefont {Jiang}\ \emph {et~al.}(2009)\citenamefont {Jiang},
  \citenamefont {Nosenko}, \citenamefont {Li}, \citenamefont {Schwabe},
  \citenamefont {Konopka}, \citenamefont {Ivlev}, \citenamefont {Fortov},
  \citenamefont {Molotkov}, \citenamefont {Lipaev}, \citenamefont {Petrov},
  \citenamefont {Turin}, \citenamefont {Thomas},\ and\ \citenamefont
  {Morfill}}]{019}%
  \BibitemOpen
  \bibfield  {author} {\bibinfo {author} {\bibfnamefont {K.}~\bibnamefont
  {Jiang}}, \bibinfo {author} {\bibfnamefont {V.}~\bibnamefont {Nosenko}},
  \bibinfo {author} {\bibfnamefont {Y.~F.}\ \bibnamefont {Li}}, \bibinfo
  {author} {\bibfnamefont {M.}~\bibnamefont {Schwabe}}, \bibinfo {author}
  {\bibfnamefont {U.}~\bibnamefont {Konopka}}, \bibinfo {author} {\bibfnamefont
  {A.~V.}\ \bibnamefont {Ivlev}}, \bibinfo {author} {\bibfnamefont {V.~E.}\
  \bibnamefont {Fortov}}, \bibinfo {author} {\bibfnamefont {V.~I.}\
  \bibnamefont {Molotkov}}, \bibinfo {author} {\bibfnamefont {A.~M.}\
  \bibnamefont {Lipaev}}, \bibinfo {author} {\bibfnamefont {O.~F.}\
  \bibnamefont {Petrov}}, \bibinfo {author} {\bibfnamefont {M.~V.}\
  \bibnamefont {Turin}}, \bibinfo {author} {\bibfnamefont {H.~M.}\ \bibnamefont
  {Thomas}}, \ and\ \bibinfo {author} {\bibfnamefont {G.~E.}\ \bibnamefont
  {Morfill}},\ }\href {\doibase 10.1209/0295-5075/85/45002} {\bibfield
  {journal} {\bibinfo  {journal} {Europhys.\ Lett.}\ }\textbf {\bibinfo
  {volume} {85}},\ \bibinfo {pages} {45002} (\bibinfo {year}
  {2009})}\BibitemShut {NoStop}%
\bibitem [{\citenamefont {Schwabe}\ \emph {et~al.}(2011)\citenamefont
  {Schwabe}, \citenamefont {Jiang}, \citenamefont {Zhdanov}, \citenamefont
  {Hagl}, \citenamefont {Huber}, \citenamefont {Ivlev}, \citenamefont {Lipaev},
  \citenamefont {Molotkov}, \citenamefont {Naumkin}, \citenamefont
  {S{\"{u}}tterlin}, \citenamefont {Thomas}, \citenamefont {Fortov},
  \citenamefont {Morfill}, \citenamefont {Skvortsov},\ and\ \citenamefont
  {Volkov}}]{19}%
  \BibitemOpen
  \bibfield  {author} {\bibinfo {author} {\bibfnamefont {M.}~\bibnamefont
  {Schwabe}}, \bibinfo {author} {\bibfnamefont {K.}~\bibnamefont {Jiang}},
  \bibinfo {author} {\bibfnamefont {S.}~\bibnamefont {Zhdanov}}, \bibinfo
  {author} {\bibfnamefont {T.}~\bibnamefont {Hagl}}, \bibinfo {author}
  {\bibfnamefont {P.}~\bibnamefont {Huber}}, \bibinfo {author} {\bibfnamefont
  {A.~V.}\ \bibnamefont {Ivlev}}, \bibinfo {author} {\bibfnamefont {A.~M.}\
  \bibnamefont {Lipaev}}, \bibinfo {author} {\bibfnamefont {V.~I.}\
  \bibnamefont {Molotkov}}, \bibinfo {author} {\bibfnamefont {V.~N.}\
  \bibnamefont {Naumkin}}, \bibinfo {author} {\bibfnamefont {K.~R.}\
  \bibnamefont {S{\"{u}}tterlin}}, \bibinfo {author} {\bibfnamefont {H.~M.}\
  \bibnamefont {Thomas}}, \bibinfo {author} {\bibfnamefont {V.~E.}\
  \bibnamefont {Fortov}}, \bibinfo {author} {\bibfnamefont {G.~E.}\
  \bibnamefont {Morfill}}, \bibinfo {author} {\bibfnamefont {A.}~\bibnamefont
  {Skvortsov}}, \ and\ \bibinfo {author} {\bibfnamefont {S.}~\bibnamefont
  {Volkov}},\ }\href {\doibase 10.1209/0295-5075/96/55001} {\bibfield
  {journal} {\bibinfo  {journal} {Europhys.\ Lett.}\ }\textbf {\bibinfo
  {volume} {96}},\ \bibinfo {pages} {55001} (\bibinfo {year}
  {2011})}\BibitemShut {NoStop}%
\bibitem [{\citenamefont {Caliebe}\ \emph {et~al.}(2011)\citenamefont
  {Caliebe}, \citenamefont {Arp},\ and\ \citenamefont {Piel}}]{11}%
  \BibitemOpen
  \bibfield  {author} {\bibinfo {author} {\bibfnamefont {D.}~\bibnamefont
  {Caliebe}}, \bibinfo {author} {\bibfnamefont {O.}~\bibnamefont {Arp}}, \ and\
  \bibinfo {author} {\bibfnamefont {A.}~\bibnamefont {Piel}},\ }\href {\doibase
  10.1063/1.3606468} {\bibfield  {journal} {\bibinfo  {journal} {Phys.\
  Plasmas}\ }\textbf {\bibinfo {volume} {18}},\ \bibinfo {pages} {073702}
  (\bibinfo {year} {2011})}\BibitemShut {NoStop}%
\bibitem [{\citenamefont {Piel}\ \emph {et~al.}(2008)\citenamefont {Piel},
  \citenamefont {Arp}, \citenamefont {Klindworth},\ and\ \citenamefont
  {Melzer}}]{12}%
  \BibitemOpen
  \bibfield  {author} {\bibinfo {author} {\bibfnamefont {A.}~\bibnamefont
  {Piel}}, \bibinfo {author} {\bibfnamefont {O.}~\bibnamefont {Arp}}, \bibinfo
  {author} {\bibfnamefont {M.}~\bibnamefont {Klindworth}}, \ and\ \bibinfo
  {author} {\bibfnamefont {A.}~\bibnamefont {Melzer}},\ }\href {\doibase
  10.1103/PhysRevE.77.026407} {\bibfield  {journal} {\bibinfo  {journal}
  {Phys.\ Rev.\ E}\ }\textbf {\bibinfo {volume} {77}},\ \bibinfo {pages}
  {026407} (\bibinfo {year} {2008})}\BibitemShut {NoStop}%
\bibitem [{\citenamefont {Menzel}\ \emph {et~al.}(2011)\citenamefont {Menzel},
  \citenamefont {Arp},\ and\ \citenamefont {Piel}}]{13}%
  \BibitemOpen
  \bibfield  {author} {\bibinfo {author} {\bibfnamefont {K.~O.}\ \bibnamefont
  {Menzel}}, \bibinfo {author} {\bibfnamefont {O.}~\bibnamefont {Arp}}, \ and\
  \bibinfo {author} {\bibfnamefont {A.}~\bibnamefont {Piel}},\ }\href {\doibase
  10.1103/PhysRevE.83.016402} {\bibfield  {journal} {\bibinfo  {journal}
  {Phys.\ Rev.\ E}\ }\textbf {\bibinfo {volume} {83}},\ \bibinfo {pages}
  {016402} (\bibinfo {year} {2011})}\BibitemShut {NoStop}%
\bibitem [{\citenamefont {Arp}\ \emph {et~al.}(2011)\citenamefont {Arp},
  \citenamefont {Caliebe},\ and\ \citenamefont {Piel}}]{14}%
  \BibitemOpen
  \bibfield  {author} {\bibinfo {author} {\bibfnamefont {O.}~\bibnamefont
  {Arp}}, \bibinfo {author} {\bibfnamefont {D.}~\bibnamefont {Caliebe}}, \ and\
  \bibinfo {author} {\bibfnamefont {A.}~\bibnamefont {Piel}},\ }\href {\doibase
  10.1103/PhysRevE.83.066404} {\bibfield  {journal} {\bibinfo  {journal}
  {Phys.\ Rev.\ E}\ }\textbf {\bibinfo {volume} {83}},\ \bibinfo {pages}
  {066404} (\bibinfo {year} {2011})}\BibitemShut {NoStop}%
\bibitem [{\citenamefont {Land}\ and\ \citenamefont {Goedheer}(2006)}]{107}%
  \BibitemOpen
  \bibfield  {author} {\bibinfo {author} {\bibfnamefont {V.}~\bibnamefont
  {Land}}\ and\ \bibinfo {author} {\bibfnamefont {W.~J.}\ \bibnamefont
  {Goedheer}},\ }\href {\doibase 10.1088/1367-2630/8/1/008} {\bibfield
  {journal} {\bibinfo  {journal} {New J.\ Phys.}\ }\textbf {\bibinfo {volume}
  {8}},\ \bibinfo {pages} {8} (\bibinfo {year} {2006})}\BibitemShut {NoStop}%
\bibitem [{\citenamefont {Feng}\ \emph {et~al.}(2016)\citenamefont {Feng},
  \citenamefont {Lin}, \citenamefont {Li},\ and\ \citenamefont {Wang}}]{106}%
  \BibitemOpen
  \bibfield  {author} {\bibinfo {author} {\bibfnamefont {Y.}~\bibnamefont
  {Feng}}, \bibinfo {author} {\bibfnamefont {W.}~\bibnamefont {Lin}}, \bibinfo
  {author} {\bibfnamefont {W.}~\bibnamefont {Li}}, \ and\ \bibinfo {author}
  {\bibfnamefont {Q.}~\bibnamefont {Wang}},\ }\href {\doibase
  10.1063/1.4962685} {\bibfield  {journal} {\bibinfo  {journal} {Phys.\
  Plasmas}\ }\textbf {\bibinfo {volume} {23}},\ \bibinfo {pages} {093705}
  (\bibinfo {year} {2016})}\BibitemShut {NoStop}%
\bibitem [{\citenamefont {Pustylnik}\ \emph {et~al.}(2017)\citenamefont
  {Pustylnik}, \citenamefont {Semenov}, \citenamefont {Zahringer},\ and\
  \citenamefont {Thomas}}]{108}%
  \BibitemOpen
  \bibfield  {author} {\bibinfo {author} {\bibfnamefont {M.~Y.}\ \bibnamefont
  {Pustylnik}}, \bibinfo {author} {\bibfnamefont {I.~L.}\ \bibnamefont
  {Semenov}}, \bibinfo {author} {\bibfnamefont {E.}~\bibnamefont {Zahringer}},
  \ and\ \bibinfo {author} {\bibfnamefont {H.~M.}\ \bibnamefont {Thomas}},\
  }\href {\doibase 10.1103/PhysRevE.96.033203} {\bibfield  {journal} {\bibinfo
  {journal} {Phys.\ Rev.\ E}\ }\textbf {\bibinfo {volume} {96}},\ \bibinfo
  {pages} {033203} (\bibinfo {year} {2017})}\BibitemShut {NoStop}%
\bibitem [{\citenamefont {Zhukhovitskii}\ \emph {et~al.}(2014)\citenamefont
  {Zhukhovitskii}, \citenamefont {Molotkov},\ and\ \citenamefont
  {Fortov}}]{22}%
  \BibitemOpen
  \bibfield  {author} {\bibinfo {author} {\bibfnamefont {D.~I.}\ \bibnamefont
  {Zhukhovitskii}}, \bibinfo {author} {\bibfnamefont {V.~I.}\ \bibnamefont
  {Molotkov}}, \ and\ \bibinfo {author} {\bibfnamefont {V.~E.}\ \bibnamefont
  {Fortov}},\ }\href {\doibase 10.1063/1.4881473} {\bibfield  {journal}
  {\bibinfo  {journal} {Phys.\ Plasmas}\ }\textbf {\bibinfo {volume} {21}},\
  \bibinfo {pages} {063701} (\bibinfo {year} {2014})}\BibitemShut {NoStop}%
\bibitem [{\citenamefont {Zhukhovitskii}\ \emph
  {et~al.}(2015{\natexlab{a}})\citenamefont {Zhukhovitskii}, \citenamefont
  {Fortov}, \citenamefont {Molotkov}, \citenamefont {Lipaev}, \citenamefont
  {Naumkin}, \citenamefont {Thomas}, \citenamefont {Ivlev}, \citenamefont
  {Schwabe},\ and\ \citenamefont {Morfill}}]{35}%
  \BibitemOpen
  \bibfield  {author} {\bibinfo {author} {\bibfnamefont {D.~I.}\ \bibnamefont
  {Zhukhovitskii}}, \bibinfo {author} {\bibfnamefont {V.~E.}\ \bibnamefont
  {Fortov}}, \bibinfo {author} {\bibfnamefont {V.~I.}\ \bibnamefont
  {Molotkov}}, \bibinfo {author} {\bibfnamefont {A.~M.}\ \bibnamefont
  {Lipaev}}, \bibinfo {author} {\bibfnamefont {V.~N.}\ \bibnamefont {Naumkin}},
  \bibinfo {author} {\bibfnamefont {H.~M.}\ \bibnamefont {Thomas}}, \bibinfo
  {author} {\bibfnamefont {A.~V.}\ \bibnamefont {Ivlev}}, \bibinfo {author}
  {\bibfnamefont {M.}~\bibnamefont {Schwabe}}, \ and\ \bibinfo {author}
  {\bibfnamefont {G.~E.}\ \bibnamefont {Morfill}},\ }\href {\doibase
  10.1063/1.4907221} {\bibfield  {journal} {\bibinfo  {journal} {Phys.\
  Plasmas}\ }\textbf {\bibinfo {volume} {22}},\ \bibinfo {pages} {023701}
  (\bibinfo {year} {2015}{\natexlab{a}})}\BibitemShut {NoStop}%
\bibitem [{\citenamefont {Zhukhovitskii}(2017)}]{86}%
  \BibitemOpen
  \bibfield  {author} {\bibinfo {author} {\bibfnamefont {D.~I.}\ \bibnamefont
  {Zhukhovitskii}},\ }\href {\doibase 10.1063/1.4978561} {\bibfield  {journal}
  {\bibinfo  {journal} {Phys.\ Plasmas}\ }\textbf {\bibinfo {volume} {24}},\
  \bibinfo {pages} {033709} (\bibinfo {year} {2017})}\BibitemShut {NoStop}%
\bibitem [{\citenamefont {Allen}(1992)}]{55}%
  \BibitemOpen
  \bibfield  {author} {\bibinfo {author} {\bibfnamefont {J.~E.}\ \bibnamefont
  {Allen}},\ }\href {\doibase 10.1088/0031-8949/45/5/013} {\bibfield  {journal}
  {\bibinfo  {journal} {Phys.\ Scr.}\ }\textbf {\bibinfo {volume} {45}},\
  \bibinfo {pages} {497} (\bibinfo {year} {1992})}\BibitemShut {NoStop}%
\bibitem [{\citenamefont {Khrapak}\ \emph {et~al.}(2012)\citenamefont
  {Khrapak}, \citenamefont {Klumov}, \citenamefont {Huber}, \citenamefont
  {Molotkov}, \citenamefont {Lipaev}, \citenamefont {Naumkin}, \citenamefont
  {Ivlev}, \citenamefont {Thomas}, \citenamefont {Schwabe}, \citenamefont
  {Morfill}, \citenamefont {Petrov}, \citenamefont {Fortov}, \citenamefont
  {Malentschenko},\ and\ \citenamefont {Volkov}}]{43}%
  \BibitemOpen
  \bibfield  {author} {\bibinfo {author} {\bibfnamefont {S.~A.}\ \bibnamefont
  {Khrapak}}, \bibinfo {author} {\bibfnamefont {B.~A.}\ \bibnamefont {Klumov}},
  \bibinfo {author} {\bibfnamefont {P.}~\bibnamefont {Huber}}, \bibinfo
  {author} {\bibfnamefont {V.~I.}\ \bibnamefont {Molotkov}}, \bibinfo {author}
  {\bibfnamefont {A.~M.}\ \bibnamefont {Lipaev}}, \bibinfo {author}
  {\bibfnamefont {V.~N.}\ \bibnamefont {Naumkin}}, \bibinfo {author}
  {\bibfnamefont {A.~V.}\ \bibnamefont {Ivlev}}, \bibinfo {author}
  {\bibfnamefont {H.~M.}\ \bibnamefont {Thomas}}, \bibinfo {author}
  {\bibfnamefont {M.}~\bibnamefont {Schwabe}}, \bibinfo {author} {\bibfnamefont
  {G.~E.}\ \bibnamefont {Morfill}}, \bibinfo {author} {\bibfnamefont {O.~F.}\
  \bibnamefont {Petrov}}, \bibinfo {author} {\bibfnamefont {V.~E.}\
  \bibnamefont {Fortov}}, \bibinfo {author} {\bibfnamefont {Y.}~\bibnamefont
  {Malentschenko}}, \ and\ \bibinfo {author} {\bibfnamefont {S.}~\bibnamefont
  {Volkov}},\ }\href {\doibase 10.1103/PhysRevE.85.066407} {\bibfield
  {journal} {\bibinfo  {journal} {Phys.\ Rev.\ E}\ }\textbf {\bibinfo {volume}
  {85}},\ \bibinfo {pages} {066407} (\bibinfo {year} {2012})}\BibitemShut
  {NoStop}%
\bibitem [{\citenamefont {S{\"{u}}tterlin}\ \emph {et~al.}(2009)\citenamefont
  {S{\"{u}}tterlin}, \citenamefont {Wysocki}, \citenamefont {Ivlev},
  \citenamefont {R{\"{a}}th}, \citenamefont {Thomas}, \citenamefont
  {Rubin-Zuzic}, \citenamefont {Goedheer}, \citenamefont {Fortov},
  \citenamefont {Lipaev}, \citenamefont {Molotkov}, \citenamefont {Petrov},
  \citenamefont {Morfill},\ and\ \citenamefont {L{\"{o}}wen}}]{68}%
  \BibitemOpen
  \bibfield  {author} {\bibinfo {author} {\bibfnamefont {K.~R.}\ \bibnamefont
  {S{\"{u}}tterlin}}, \bibinfo {author} {\bibfnamefont {A.}~\bibnamefont
  {Wysocki}}, \bibinfo {author} {\bibfnamefont {A.~V.}\ \bibnamefont {Ivlev}},
  \bibinfo {author} {\bibfnamefont {C.}~\bibnamefont {R{\"{a}}th}}, \bibinfo
  {author} {\bibfnamefont {H.~M.}\ \bibnamefont {Thomas}}, \bibinfo {author}
  {\bibfnamefont {M.}~\bibnamefont {Rubin-Zuzic}}, \bibinfo {author}
  {\bibfnamefont {W.~J.}\ \bibnamefont {Goedheer}}, \bibinfo {author}
  {\bibfnamefont {V.~E.}\ \bibnamefont {Fortov}}, \bibinfo {author}
  {\bibfnamefont {A.~M.}\ \bibnamefont {Lipaev}}, \bibinfo {author}
  {\bibfnamefont {V.~I.}\ \bibnamefont {Molotkov}}, \bibinfo {author}
  {\bibfnamefont {O.~F.}\ \bibnamefont {Petrov}}, \bibinfo {author}
  {\bibfnamefont {G.~E.}\ \bibnamefont {Morfill}}, \ and\ \bibinfo {author}
  {\bibfnamefont {H.}~\bibnamefont {L{\"{o}}wen}},\ }\href {\doibase
  10.1103/PhysRevLett.102.085003} {\bibfield  {journal} {\bibinfo  {journal}
  {Phys.\ Rev.\ Lett.}\ }\textbf {\bibinfo {volume} {102}},\ \bibinfo {pages}
  {085003} (\bibinfo {year} {2009})}\BibitemShut {NoStop}%
\bibitem [{\citenamefont {Morfill}\ \emph {et~al.}(2012)\citenamefont
  {Morfill}, \citenamefont {Ivlev},\ and\ \citenamefont {Thomas}}]{63}%
  \BibitemOpen
  \bibfield  {author} {\bibinfo {author} {\bibfnamefont {G.~E.}\ \bibnamefont
  {Morfill}}, \bibinfo {author} {\bibfnamefont {A.~V.}\ \bibnamefont {Ivlev}},
  \ and\ \bibinfo {author} {\bibfnamefont {H.~M.}\ \bibnamefont {Thomas}},\
  }\href {\doibase 10.1063/1.4717979} {\bibfield  {journal} {\bibinfo
  {journal} {Phys.\ Plasmas}\ }\textbf {\bibinfo {volume} {19}},\ \bibinfo
  {pages} {055402} (\bibinfo {year} {2012})}\BibitemShut {NoStop}%
\bibitem [{\citenamefont {Khrapak}\ \emph {et~al.}(2016)\citenamefont
  {Khrapak}, \citenamefont {Molotkov}, \citenamefont {Lipaev}, \citenamefont
  {Zhukhovitskii}, \citenamefont {Naumkin}, \citenamefont {Fortov},
  \citenamefont {Petrov}, \citenamefont {Thomas}, \citenamefont {Khrapak},
  \citenamefont {Huber}, \citenamefont {Ivlev},\ and\ \citenamefont
  {Morfill}}]{92}%
  \BibitemOpen
  \bibfield  {author} {\bibinfo {author} {\bibfnamefont {A.~G.}\ \bibnamefont
  {Khrapak}}, \bibinfo {author} {\bibfnamefont {V.~I.}\ \bibnamefont
  {Molotkov}}, \bibinfo {author} {\bibfnamefont {A.~M.}\ \bibnamefont
  {Lipaev}}, \bibinfo {author} {\bibfnamefont {D.~I.}\ \bibnamefont
  {Zhukhovitskii}}, \bibinfo {author} {\bibfnamefont {V.~N.}\ \bibnamefont
  {Naumkin}}, \bibinfo {author} {\bibfnamefont {V.~E.}\ \bibnamefont {Fortov}},
  \bibinfo {author} {\bibfnamefont {O.~F.}\ \bibnamefont {Petrov}}, \bibinfo
  {author} {\bibfnamefont {H.~M.}\ \bibnamefont {Thomas}}, \bibinfo {author}
  {\bibfnamefont {S.~A.}\ \bibnamefont {Khrapak}}, \bibinfo {author}
  {\bibfnamefont {P.}~\bibnamefont {Huber}}, \bibinfo {author} {\bibfnamefont
  {A.}~\bibnamefont {Ivlev}}, \ and\ \bibinfo {author} {\bibfnamefont
  {G.}~\bibnamefont {Morfill}},\ }\href {\doibase 10.1002/ctpp.201500102}
  {\bibfield  {journal} {\bibinfo  {journal} {Contrib.\ Plasma Phys.}\ }\textbf
  {\bibinfo {volume} {56}},\ \bibinfo {pages} {253} (\bibinfo {year}
  {2016})}\BibitemShut {NoStop}%
\bibitem [{\citenamefont {Knapek}\ \emph {et~al.}(2018)\citenamefont {Knapek},
  \citenamefont {Huber}, \citenamefont {Mohr}, \citenamefont {Zaehringer},
  \citenamefont {Molotkov}, \citenamefont {Lipaev}, \citenamefont {Naumkin},
  \citenamefont {Konopka}, \citenamefont {Thomas},\ and\ \citenamefont
  {Fortov}}]{110}%
  \BibitemOpen
  \bibfield  {author} {\bibinfo {author} {\bibfnamefont {C.~A.}\ \bibnamefont
  {Knapek}}, \bibinfo {author} {\bibfnamefont {P.}~\bibnamefont {Huber}},
  \bibinfo {author} {\bibfnamefont {D.~P.}\ \bibnamefont {Mohr}}, \bibinfo
  {author} {\bibfnamefont {E.}~\bibnamefont {Zaehringer}}, \bibinfo {author}
  {\bibfnamefont {V.~I.}\ \bibnamefont {Molotkov}}, \bibinfo {author}
  {\bibfnamefont {A.~M.}\ \bibnamefont {Lipaev}}, \bibinfo {author}
  {\bibfnamefont {V.}~\bibnamefont {Naumkin}}, \bibinfo {author} {\bibfnamefont
  {U.}~\bibnamefont {Konopka}}, \bibinfo {author} {\bibfnamefont {H.~M.}\
  \bibnamefont {Thomas}}, \ and\ \bibinfo {author} {\bibfnamefont {V.~E.}\
  \bibnamefont {Fortov}},\ }\href {\doibase 10.1063/1.5020392} {\bibfield
  {journal} {\bibinfo  {journal} {AIP Conf.\ Proc.}\ }\textbf {\bibinfo
  {volume} {1925}},\ \bibinfo {pages} {020004} (\bibinfo {year}
  {2018})}\BibitemShut {NoStop}%
\bibitem [{\citenamefont {Zhukhovitskii}\ \emph {et~al.}(1984)\citenamefont
  {Zhukhovitskii}, \citenamefont {Khrapak},\ and\ \citenamefont
  {Yakubov}}]{103}%
  \BibitemOpen
  \bibfield  {author} {\bibinfo {author} {\bibfnamefont {D.~I.}\ \bibnamefont
  {Zhukhovitskii}}, \bibinfo {author} {\bibfnamefont {A.~G.}\ \bibnamefont
  {Khrapak}}, \ and\ \bibinfo {author} {\bibfnamefont {I.~T.}\ \bibnamefont
  {Yakubov}},\ }\href@noop {} {\bibfield  {journal} {\bibinfo  {journal}
  {Teplofiz.\ Vys.\ Temp.\ (High Temperature)}\ }\textbf {\bibinfo {volume}
  {22}},\ \bibinfo {pages} {833} (\bibinfo {year} {1984})}\BibitemShut
  {NoStop}%
\bibitem [{\citenamefont {Filippov}\ \emph {et~al.}(2007)\citenamefont
  {Filippov}, \citenamefont {Zagorodny}, \citenamefont {Momot}, \citenamefont
  {Pal'},\ and\ \citenamefont {Starostin}}]{104}%
  \BibitemOpen
  \bibfield  {author} {\bibinfo {author} {\bibfnamefont {A.~V.}\ \bibnamefont
  {Filippov}}, \bibinfo {author} {\bibfnamefont {A.~G.}\ \bibnamefont
  {Zagorodny}}, \bibinfo {author} {\bibfnamefont {A.~I.}\ \bibnamefont
  {Momot}}, \bibinfo {author} {\bibfnamefont {A.~F.}\ \bibnamefont {Pal'}}, \
  and\ \bibinfo {author} {\bibfnamefont {A.~N.}\ \bibnamefont {Starostin}},\
  }\href@noop {} {\bibfield  {journal} {\bibinfo  {journal} {J.\ Exp.\ Theor.\
  Phys.}\ }\textbf {\bibinfo {volume} {104}},\ \bibinfo {pages} {147} (\bibinfo
  {year} {2007})}\BibitemShut {NoStop}%
\bibitem [{\citenamefont {Zhukhovitskii}\ \emph {et~al.}(2017)\citenamefont
  {Zhukhovitskii}, \citenamefont {Naumkin}, \citenamefont {Khusnulgatin},
  \citenamefont {Molotkov},\ and\ \citenamefont {Lipaev}}]{94}%
  \BibitemOpen
  \bibfield  {author} {\bibinfo {author} {\bibfnamefont {D.~I.}\ \bibnamefont
  {Zhukhovitskii}}, \bibinfo {author} {\bibfnamefont {V.~N.}\ \bibnamefont
  {Naumkin}}, \bibinfo {author} {\bibfnamefont {A.~I.}\ \bibnamefont
  {Khusnulgatin}}, \bibinfo {author} {\bibfnamefont {V.~I.}\ \bibnamefont
  {Molotkov}}, \ and\ \bibinfo {author} {\bibfnamefont {A.~M.}\ \bibnamefont
  {Lipaev}},\ }\href {\doibase 10.1103/PhysRevE.96.043204} {\bibfield
  {journal} {\bibinfo  {journal} {Phys.\ Rev.\ E}\ }\textbf {\bibinfo {volume}
  {96}},\ \bibinfo {pages} {043204} (\bibinfo {year} {2017})}\BibitemShut
  {NoStop}%
\bibitem [{\citenamefont {Zhukhovitskii}\ \emph
  {et~al.}(2015{\natexlab{b}})\citenamefont {Zhukhovitskii}, \citenamefont
  {Petrov}, \citenamefont {Hyde}, \citenamefont {Herdrich}, \citenamefont
  {Laufer}, \citenamefont {Dropmann},\ and\ \citenamefont {Matthews}}]{47}%
  \BibitemOpen
  \bibfield  {author} {\bibinfo {author} {\bibfnamefont {D.~I.}\ \bibnamefont
  {Zhukhovitskii}}, \bibinfo {author} {\bibfnamefont {O.~F.}\ \bibnamefont
  {Petrov}}, \bibinfo {author} {\bibfnamefont {T.~W.}\ \bibnamefont {Hyde}},
  \bibinfo {author} {\bibfnamefont {G.}~\bibnamefont {Herdrich}}, \bibinfo
  {author} {\bibfnamefont {R.}~\bibnamefont {Laufer}}, \bibinfo {author}
  {\bibfnamefont {M.}~\bibnamefont {Dropmann}}, \ and\ \bibinfo {author}
  {\bibfnamefont {L.}~\bibnamefont {Matthews}},\ }\href {\doibase
  10.1088/1367-2630/17/5/053041} {\bibfield  {journal} {\bibinfo  {journal}
  {New J.\ Phys.}\ }\textbf {\bibinfo {volume} {17}},\ \bibinfo {pages}
  {053041} (\bibinfo {year} {2015}{\natexlab{b}})}\BibitemShut {NoStop}%
\bibitem [{\citenamefont {Lampe}\ \emph {et~al.}(2003)\citenamefont {Lampe},
  \citenamefont {Goswami}, \citenamefont {Sternovsky}, \citenamefont
  {Robertson}, \citenamefont {Gavrishchaka}, \citenamefont {Ganguli},\ and\
  \citenamefont {Joyce}}]{105}%
  \BibitemOpen
  \bibfield  {author} {\bibinfo {author} {\bibfnamefont {M.}~\bibnamefont
  {Lampe}}, \bibinfo {author} {\bibfnamefont {R.}~\bibnamefont {Goswami}},
  \bibinfo {author} {\bibfnamefont {Z.}~\bibnamefont {Sternovsky}}, \bibinfo
  {author} {\bibfnamefont {S.}~\bibnamefont {Robertson}}, \bibinfo {author}
  {\bibfnamefont {V.}~\bibnamefont {Gavrishchaka}}, \bibinfo {author}
  {\bibfnamefont {G.}~\bibnamefont {Ganguli}}, \ and\ \bibinfo {author}
  {\bibfnamefont {G.}~\bibnamefont {Joyce}},\ }\href {\doibase
  10.1063/1.1562163} {\bibfield  {journal} {\bibinfo  {journal} {Phys.\
  Plasmas}\ }\textbf {\bibinfo {volume} {10}},\ \bibinfo {pages} {1500}
  (\bibinfo {year} {2003})}\BibitemShut {NoStop}%
\bibitem [{\citenamefont {Zobnin}\ \emph {et~al.}(2000)\citenamefont {Zobnin},
  \citenamefont {Nefedov}, \citenamefont {Sinel'shchikov},\ and\ \citenamefont
  {Fortov}}]{67}%
  \BibitemOpen
  \bibfield  {author} {\bibinfo {author} {\bibfnamefont {A.~V.}\ \bibnamefont
  {Zobnin}}, \bibinfo {author} {\bibfnamefont {A.~P.}\ \bibnamefont {Nefedov}},
  \bibinfo {author} {\bibfnamefont {V.~A.}\ \bibnamefont {Sinel'shchikov}}, \
  and\ \bibinfo {author} {\bibfnamefont {V.~E.}\ \bibnamefont {Fortov}},\
  }\href {\doibase 10.1134/1.1320081} {\bibfield  {journal} {\bibinfo
  {journal} {J.\ Exp.\ Theor.\ Phys.}\ }\textbf {\bibinfo {volume} {91}},\
  \bibinfo {pages} {483} (\bibinfo {year} {2000})}\BibitemShut {NoStop}%
\bibitem [{\citenamefont {Zhukhovitskii}(2015)}]{64}%
  \BibitemOpen
  \bibfield  {author} {\bibinfo {author} {\bibfnamefont {D.~I.}\ \bibnamefont
  {Zhukhovitskii}},\ }\href {\doibase 10.1103/PhysRevE.92.023108} {\bibfield
  {journal} {\bibinfo  {journal} {Phys.\ Rev.\ E}\ }\textbf {\bibinfo {volume}
  {92}},\ \bibinfo {pages} {023108} (\bibinfo {year} {2015})}\BibitemShut
  {NoStop}%
\bibitem [{\citenamefont {Naumkin}\ \emph {et~al.}(2016)\citenamefont
  {Naumkin}, \citenamefont {Zhukhovitskii}, \citenamefont {Molotkov},
  \citenamefont {Lipaev}, \citenamefont {Fortov}, \citenamefont {Thomas},
  \citenamefont {Huber},\ and\ \citenamefont {Morfill}}]{69}%
  \BibitemOpen
  \bibfield  {author} {\bibinfo {author} {\bibfnamefont {V.~N.}\ \bibnamefont
  {Naumkin}}, \bibinfo {author} {\bibfnamefont {D.~I.}\ \bibnamefont
  {Zhukhovitskii}}, \bibinfo {author} {\bibfnamefont {V.~I.}\ \bibnamefont
  {Molotkov}}, \bibinfo {author} {\bibfnamefont {A.~M.}\ \bibnamefont
  {Lipaev}}, \bibinfo {author} {\bibfnamefont {V.~E.}\ \bibnamefont {Fortov}},
  \bibinfo {author} {\bibfnamefont {H.~M.}\ \bibnamefont {Thomas}}, \bibinfo
  {author} {\bibfnamefont {P.}~\bibnamefont {Huber}}, \ and\ \bibinfo {author}
  {\bibfnamefont {G.~E.}\ \bibnamefont {Morfill}},\ }\href {\doibase
  10.1103/PhysRevE.94.033204} {\bibfield  {journal} {\bibinfo  {journal}
  {Phys.\ Rev.\ E}\ }\textbf {\bibinfo {volume} {94}},\ \bibinfo {pages}
  {033204} (\bibinfo {year} {2016})}\BibitemShut {NoStop}%
\bibitem [{\citenamefont {Saitou}(2018)}]{109}%
  \BibitemOpen
  \bibfield  {author} {\bibinfo {author} {\bibfnamefont {Y.}~\bibnamefont
  {Saitou}},\ }\href {\doibase 10.1063/1.5037020} {\bibfield  {journal}
  {\bibinfo  {journal} {Phys.\ Plasmas}\ }\textbf {\bibinfo {volume} {25}},\
  \bibinfo {pages} {073701} (\bibinfo {year} {2018})}\BibitemShut {NoStop}%
\end{thebibliography}
\end{document}